\input harvmac 

\parskip=4pt \baselineskip=12pt
\parindent 10pt


\def\nl{\hfil\break} 
\def\nt{\noindent} 
\hfuzz=18pt 
\font\male=cmr9 
\font\fmale=cmr8 
\font\sfont=cmbx10 scaled\magstep1

\def\newsubsec#1{\global\advance\subsecno
by1\message{(\secsym\the\subsecno. #1)}
\ifnum\lastpenalty>9000\else\bigbreak\fi
\noindent{\bf\secsym\the\subsecno. #1}\writetoca{\string\quad
{\secsym\the\subsecno.} {#1}}}

\def\dia{{$\diamondsuit$}}
\def\han{{\textstyle{n\over2}}}
\def\has{{\textstyle{n\over6}}}
\def\thf{{\textstyle{3\over2}}} 
\def\tn{{\tilde N}}
\def\nn{{\tilde n}}

\def\bu{{$\bullet$}} 
\font\verysmall=cmr5
\def\phr{\raise1.1pt\hbox{\verysmall x}\kern-8pt\supset}

\def\bhr{\raise1.1pt\hbox{\verysmall x}\kern-9pt\subset}

\def\llra{\longleftrightarrow} 
\def\bac{{C\kern-5.2pt I}} \def\bbc{{C\kern-6.5pt I}} 
\def\mt{\mapsto} \def\D{\Delta} \def\bbz{Z\!\!\!Z}
\def\bbr{I\!\!R} \def\bbn{I\!\!N}
\def\a{\alpha} \def\b{\beta} \def\g{\gamma} \def\d{\delta}

\def\L{\Lambda} \def\hL{{\hat\Lambda}}
 
\def\cd{{\cal D}}  \def\cb{{\cal B}} 
\def\cc{{\cal C}} \def\cih{{\cal C}_\chi} 

\def\tcih{{\tilde{\cal C}}_\chi}
\def\tcihp{{\tilde{\cal C}}_{\chi'}}

\def\tih{{\tilde{\cal T}}^\chi}
\def\tihp{{\tilde{\cal T}}^{\chi'}} 

 \def\ct{{\cal T}} \def\cf{{\cal F}}
\def\ch{{\cal H}}  \def\cl{{\cal L}}
\def\ct{{\cal T}} \def\cs{{\cal S}} 
 \def\cn{{\cal N}} \def\tcn{{\tilde{\cal N}}}
\def\ca{{\cal A}}
 \def\cp{{\cal P}}  
\def\cg{{\cal G}} \def\gc{{\cal G}_c} 
\def\tcl{{\tilde{\cal L}}}

\def\r{\rho} \def\eps{\epsilon} \def\s{\sigma}

\def\lg{\langle} \def\rg{\rangle} 
\def\vf{\varphi}  \def\G{\Gamma} 
  
\def\lra{\longrightarrow} \def\Lra{\Longrightarrow} 
 \def\vr{\vert}
\def\bl{{\bar \ell}}

 \def\ci{{\cal I}} 
  
\def\pd{\partial}   
 \def\ha{{\hat a}}

\def\nl{\hfil\break} \def\np{\vfil\eject}
\def\ni{\noindent} \def\st{\vskip 3mm\ni}
\def\ho{\hat\otimes} 
\def\ido{intertwining differential operator}
\def\idos{intertwining differential operators} 
 


\nref\Baa{Baston R.J.: Verma modules and differential conformal
invariants. J. Diff. Geom. {\bf 32}, 851-898 (1990) }

\nref\Bab{Baston R.J.: Almost Hermitian symmetric manifolds, I:
Local twistor theory, II: Differential invariants. Duke. Math.
J. {\bf 63}, 81-111, 113-138 (1991) }

\nref\BEG{Bailey T.N., Eastwood M.G., Graham C.R.: Invariant
theory for conformal and CR geometry. Ann. Math. {\bf 139},
491-552 (1994) }

\nref\BGG{Bernstein I.N., Gel'fand I.M., Gel'fand S.I.: Structure
of representations generated by highest weight vectors. Funkts.
Anal. Prilozh. {\bf 5} (1), 1--9 (1971); English translation:
Funkt. Anal. Appl. {\bf 5}, 1--8 (1971)}

\nref\BC{ Boe B.D., Collingwood D.H.: A comparison theory for the
structure of induced representations I \& II. J. Algebra, {\bf
94}, 511-545 (1985) \& Math. Z. {\bf 190}, 1-11 (1085) }

\nref\Br{ Branson T.P.: Differential operators canonically
associated to a conformal structure. Math. Scand. {\bf 57},
293-345 (1985) }

\nref\CSS{\v Cap A., Slov\'ak J., Sou\v cek V.: Invariant
operators on manifolds with almost hermitian structures, I.
Invariant differentiation. preprint ESI 186, Vienna, 1994: II.
Normal Cartan connections. to appear}

\nref\Di{ Dixmier J.: Enveloping Algebras. New York: North
Holland 1977}

\nref\Doj{Dobrev V.K.: Elementary representations and
intertwining operators for $SU(2,2)$: I. J. Math. Phys. {\bf 26},
235-251 (1985)}

\nref\Dob{Dobrev V.K.: Canonical construction of intertwining
differential operators associated with representations of real
semisimple Lie groups. Rep. Math. Phys. {\bf 25}, 159-181 (1988)
}

\nref\DMPPT{Dobrev V.K., Mack G., Petkova V.B., Petrova S.G.,
Todorov I.T.: Harmonic Analysis on the $n$ - Dimensional
Lorentz Group and Its Applications to Conformal Quantum Field
Theory. Lecture Notes in Phys., No 63, Berlin: Springer-Verlag
1977}

\nref\DP{Dobrev V.K., Petkova V.B.: Elementary representations
and intertwining operators for the group $SU^*(4)$. Rep. Math.
Phys. {\bf 13}, 233-277 (1978) }

\nref\EG{Eastwood M.G., Graham, C.R.: Invariants of conformal
densities, Duke. Math. J. {\bf 63}, 633-671 (1991) }

\nref\ER{ Eastwood M.G., Rice J.W.: Conformally invariant
differential operators on Minkowski space and their curved
analogues. Commun. Math. Phys. {\bf 109}, 207-228 (1987) }

\nref\GGV{Gel'fand I.M., Graev M.I., Vilenkin N.Ya.: Generalized
Functions, Vol. 5: Integral Geometry and Representation Theory.
New York Academic Press 1966}

\nref\Goa{ Gover A.R.: Conformally invariant operators of
standard type. Quart. J. Math. {\bf 40}, 197-208 (1989) }

\nref\Ja{Jakobsen H.P.: A spin-off from highest weight
representations; conformal covariants, in particular for
$O(3,2)$. in: Lecture Notes in Phys. Vol. 261, pp. 253-265.
Berlin: Springer-Verlag 1986; \nl Conformal covariants. Publ.
RIMS, Kyoto Univ. {\bf 22}, 345-364 (1986) }

\nref\JV{Jakobsen H.P., Vergne M.: Wave and Dirac operators, and
representations of the conformal group. J. Funct. Anal. {\bf
24}, 52-106 (1977)}

\nref\Kia{Kirillov A.A.: Orbits of the group of diffeomorphisms
of a circle and local Lie superalgebras, Funkts. Anal. Prilozh.
{\bf 15} (2), 75-76 (1981);\ English translation: Funct. Anal.
Appl. {\bf 15}, 135-137 (1981)}

\nref\Kib{Kirillov A.A.: Infinite dimensional Lie groups, their
orbits, invariants and representations. The geometry of moments.
In: Doebner H.-D., Palev T.D. (eds.) Twistor Geometry and
Non-Linear Systems. Proceedings, 4th Bulgarian Summer School on
Mathematical Problems of Quantum Field Theory, Primorsko, 1980,
pp. 101-123. Lecture Notes in Math., Vol. 970. Berlin:
Springer-Verlag 1982}

\nref\Kn{Knapp A.W.: Representation Theory of Semisimple Groups
(An Overview Based on Examples). Princeton: Princeton Univ. Press
1986}

\nref\KS{Knapp A.W., Stein E.M.: Intertwining operators for
semisimple groups. Ann. Math. {\bf 93}, 489-578 (1971); ~II:
Inv. Math. {\bf 60}, 9-84 (1980) }

\nref\KZ{Knapp A.W., Zuckerman G.J.: Classification theorems for
representations of semisimple Lie groups. In: Lecture Notes in
Math. Vol. 587, pp. 138-159. Berlin: Springer-Verlag 1977; \nl
Classification of irreducible tempered representations of
semisimple groups. Ann. Math. {\bf 116}, 389-501 (1982) }

\nref\Ko{ Kostant B.: Verma modules and the existence of
quasi-invariant differential operators. In: Lecture Notes in
Math., Vol. 466. p. 101. Berlin: Springer-Verlag 1975}

\nref\La{ Langlands R.P.: On the classification of irreducible
representations of real algebraic groups. preprint, 
 Institute for Advanced Study, Princeton (1973); 
published in: Representation Theory and Harmonic Analysis 
on Semisimple Lie Groups, eds. P. Sally and D. Vogan 
(1989) pp. 101-170.}

\nref\Oe{ \O rsted B.: Conformally invariant differential
equations and projective geometry. J. Funct. Anal. {\bf 44}, 1-23
(1981) }

\nref\PS{Petkova V.B., Sotkov G.M.: The six-point families of
exceptional representations of the conformal group. Lett. Math.
Phys. {\bf 8}, 217-226 (1984) }

\nref\SV{ Speh B., Vogan D.A.: Reducibility of generalized
principal series representations. Acta Math. {\bf 145}, 227-299
(1980) }

\lref\KNV{Volkov A.Yu., {\it Lett. Math. Phys.} (1997) {\bf 39}, 313.} 

\nref\Wey{ Weyl H.: The Classical Groups. Princeton: Princeton
Univ. Press, 1939}

\nref\Wu{ W\"unsch V.: On conformally invariant differential
operators. Math. Nachr. {\bf 129}, 269-281 (1986) }

\nref\Zha{ Zhelobenko D.P.: Harmonic Analysis on Semisimple
Complex Lie Groups. Moscow: Nauka 1974 (in Russian)}

\nref\Zhb{ Zhelobenko D.P.: Discrete symmetry operators for
reductive Lie groups. Math. USSR Izv. {\bf 40}, 1055-1083 (1976)
}


\parskip 3pt
\baselineskip=12pt 
\line{\hfill J. Geom. Phys. {\bf 25} (1998) 1-28 
 (first as preprint ASI-TPA/11/95 (April 1995))} 

\vskip 1truecm

\centerline{{\sfont New Generalized Verma Modules and}}
\vskip 2mm 
\centerline{{\sfont Multilinear Intertwining Differential
Operators}} 

\footnote{}{ ~\fmale{Key words: generalized Verma modules,
intertwining operators, multilinear differential operators;
~~MSC~: ~~81R05, 22E47, 17B35, 34G20}}

\vskip 1.5cm 

\centerline{{\bf V.K. Dobrev}$^{*}$} \footnote{}{$^{*}$ 
~\male{Permanent address : Bulgarian Academy of Sciences,
Institute of Nuclear Research and Nuclear Energy, 72
Tsarigradsko Chaussee, 1784 Sofia, Bulgaria;
e-mail: dobrev@inrne.bas.bg}} 

\vskip 3mm

\centerline{Arnold Sommerfeld Institute for Mathematical Physics}
\centerline{Technical University Clausthal} 
\centerline{Leibnizstr. 10, 38678 Clausthal-Zellerfeld, Germany}
\centerline{e-mail: ptvd@pt.tu-clausthal.de} 

\vskip 1mm 

\centerline{and}

\vskip 1mm

\centerline{
International Center for Theoretical Physics} 
\centerline{via Costiera 11, P.O. Box 586} 
\centerline{34100 Trieste, Italy}
\centerline{e-mail: dobrev@ictp.trieste.it} 

\vskip 1cm

\centerline{\bf Abstract}

\midinsert\male\noindent 
The present paper contains ~{\it two}~ interrelated developments.
~{\it First}~ are proposed new generalized Verma modules. They
are called ~$k$ - Verma modules, ~$k\in\bbn$, and coincide with
the usual Verma modules for $k=1$. As a vector space a ~$k$ -
Verma module is isomorphic to the symmetric tensor product of
~$k$~ copies of the universal enveloping algebra ~$U(\cg^-)$,
where ~$\cg^-$~ is the subalgebra of lowering generators in the
standard triangular decomposition of a simple Lie algebra ~$\cg
~=~ \cg^+ \oplus \ch \oplus \cg^-$. The ~{\it second}~
development is the proposal of a procedure for the construction
of ~{\it multilinear} \idos\ for semisimple Lie groups ~$G$. This
procedure uses ~$k$ - Verma modules and coincides for ~$k=1$~
with a procedure for the construction of ~{\it linear} \idos. For
all ~$k$~ central role is played by the singular vectors of the
~$k$ - Verma modules. Explicit formulae for series of such
singular vectors are given. Using these are given explicitly many
new examples of multilinear \idos. In particular, for ~$G ~=~
SL(2,\bbr)$~ are given explicitly all bilinear \idos. Using the
latter, as an application are constructed ~$(n/2)$-differentials
for all ~$n\in 2\bbn$, the ordinary Schwarzian being the case
$n=4$. As another application, in a Note Added we propose a new 
hierarchy of nonlinear equations, the lowest member being the 
KdV equation. 
\endinsert

\np 

\baselineskip=12pt plus 2pt minus 1pt 
\parskip=7pt plus 1pt

\newsec{Introduction}

\newsubsec{} 
Operators intertwining representations of Lie groups play a very
important role both in mathematics and physics. To recall the
notions, consider a Lie group ~$G$~ and two representations
~$T,T'$~ of ~$G$~ acting in the representation spaces ~$C,C'$,
which may be Hilbert, Fr\'echet, etc. An ~{\it intertwining
operator} ~$\ci$~ for these two representations is a continuous
linear map
\eqn\ima{ \ci ~:~ C \lra C', \quad \ci ~:~ f ~ \mt ~ j, \quad 
 \quad f\in C, ~j\in C' } 
such that 
\eqn\int{ \ci \circ T(g) ~~=~~ T'(g) \circ \ci ~, \quad \forall g\in
G }
An important application of the intertwining operators is that
they produce canonically invariant equations. Indeed, in the
setting above the equation
\eqn\inv{ \ci~f ~~=~~ j } 
is a ~$G$-invariant equation. These are very useful in the
applications, recall, e.g., the well known examples of Dirac,
Maxwell equations. The intertwining operators are also very
relevant for analyzing the structure of representations of Lie
groups, especially of semisimple (or reductive) Lie groups, cf.,
e.g., \KS, \Ko, \Zha, \Zhb. There are two types of intertwining
operators : integral and differential. For the integral
intertwining operators, which we shall not discuss here, we refer
to \KS, \Zha\ for the mathematical side and to \DMPPT\ for
explicit examples and applications. For the \idos\ we refer to
\Ko, \Zhb, \Dob; (for early examples and partial cases see, e.g.,
\DMPPT, \JV, \DP, \Oe, \PS, \Doj, \Ja, \BC, \Br, \Wu, \ER, \Goa,
\Baa, \Bab). 

\newsubsec{} 
In the present paper we discuss ~{\it multilinear \idos}~ such
that:
\eqn\imb{ _k\ci ~~:~~ \underbrace{ f \otimes \dots \otimes f }_k 
 ~ \mt ~ j ~, \quad f \in C, ~j\in C' }
\eqn\intb{ _k\ci ~\circ~
\underbrace{ T(g) \otimes \dots \otimes T(g) }_k 
 ~~=~~ T'(g) ~\circ ~_k\ci ~, \quad \forall g\in G }
Clearly, for ~$k=1$~ \imb, \intb\ reduce to \ima, \int, resp. 

Let us give an example of such operator for ~$k=2$. Let ~$G ~=~
SL(2,\bbr)$~ and consider ~$C^\infty$ - functions so that the
representation is acting as \GGV~:
\eqna\rpr 
$$\eqalignno{ 
&T^c(g) ~ f(x) ~~=~~ \vert \d - \b x \vert^{-c}~ 
f \left( {{ \a x -\g \over \d - \b x}} \right)~, \quad 
\d - \b x \neq 0 &\rpr a\cr 
&g ~=~ \pmatrix{ \a & \b \cr \g &\d}, ~~\a\d-\b\g=1,
~\a,\b,\g,d \in\bbr &\rpr b\cr 
}$$
where ~$c\in\bbc$~ is a parameter characterizing the
representation (for more details see Section 4). Consider now
the operator:
\eqn\schw{ _2\ci \left( f \right) ~~=~~f'''~f' ~-~ \thf ~(f'')^2
} 
where ~$f',f'',f'''$~ are the first, second, third, resp.
derivatives of ~$f$. Let us denote the space of functions with
\rpr{} as transformation rule by ~$\cc^c$. Then it is easy to
show that ~$_2\ci$~ has the following intertwining property:
\eqn\imc{ _2\ci ~~:~~ f ~\otimes ~f 
 ~ \mt ~ j ~, \quad f \in \cc^0 \,, ~~j \in \cc^8 } 
\eqn\intb{ _2\ci ~\circ~ \left( 
T^0(g) \otimes T^0(g) \right) 
~~=~~ T^8(g) ~\circ ~_2\ci ~, \quad \forall g\in G }

\newsubsec{} 
We would like to note that our problem is related to the problem
of finding invariant ~$n$--differentials. In the simple example
above such a relation is straightforward. Indeed an example of
invariant quadratic differential is the Schwarzian (cf., e.g.,
\Kib)~:
\eqn\schwa{\eqalign{ 
&{\rm Sch} (f) ~~\doteq~~ \left( {f'''\over f'} ~-~ \thf ~{(f'')^2
\over (f')^2 }\right) ~(dx)^2 \cr
&{\rm Sch} \left( f \circ f_0 
\right) ~~=~~ 
{\rm Sch} \left(f(x)\right) ~\circ ~ f_0 \,, \quad 
f_0(x) ~=~ { \a x -\g \over \d - \b x} 
\cr
&{\rm Sch} (f_0) ~~=~~ 0 
}} 

Of course, such direct relations are an artifact of the
simplicity of the situation. Our setting is more general than the
problem of finding invariants as in \schwa\ since it allows in
principle arbitrary representation parameters. Below we give such
operators for any semisimple Lie group.

For other examples of multilinear invariant operators see, e.g.,
\EG, \BEG. These examples rely on adaptations of the classical
polynomial invariant theory of Weyl \Wey. Another approach is to
use invariant differentiation with respect to a Cartan connection
\CSS.

\newsubsec{} 
{\it Our approach is different}\ from those just mentioned. It is
a natural generalization of the ~$k=1$~ procedure of \Dob. More
than that. The present paper contains ~{\bf two}~ interrelated
developments. ~{\bf First}~ we propose ~{\it new generalized
Verma modules}. They are called ~$k$ - Verma modules,
~$k\in\bbn$, and coincide with the usual Verma modules for $k=1$.
As a vector space a ~$k$ - Verma module is isomorphic to the
symmetric tensor product of ~$k$~ copies of the universal
enveloping algebra ~$U(\cg^-)$, where ~$\cg^-$~ is the subalgebra
of lowering generators in the standard triangular decomposition
of a simple Lie algebra ~$\cg ~=~ \cg^+ \oplus \ch \oplus \cg^-$.
The ~{\bf second}~ development is the proposal of a procedure for
the construction of ~{\it multilinear \idos}\ for semisimple Lie
groups ~$G$. This procedure uses the ~$k$ - Verma modules and
coincides for ~$k=1$~ with our procedure for the construction of
~{\it linear} \idos\ \Dob. For all ~$k$~ central role is played
by the singular vectors of the ~$k$ - Verma modules. Explicit
formulae for series of such singular vectors are given for
arbitrary $\cg$. Using these are given explicitly many new
examples of multilinear \idos. In particular, for ~$G ~=~
SL(2,\bbr)$~ are given explicitly all bilinear \idos. Using the
latter, as an application are constructed ~$(n/2)$-differentials
for all ~$n\in 2\bbn$, the ordinary Schwarzian being the case
$n=4$.

\newsubsec{} 
The organization of the paper is as follows. 

In Section 2 we first recall the usual Verma modules formulating
their reducibility conditions in a way suitable for our purposes.
Then we introduce the new generalization of the Verma modules,
which we call ~$k$ - Verma modules, and we give some of their
general properties.

In Section 2 we consider the singular vectors of the $k$-Verma
modules. Using the singular vectors we show that $k$-Verma
modules are always reducible independently of the highest weight
in sharp contrast with the ordinary Verma modules ($k=1$). We
also give many important explicit examples of singular vectors
for $k=2,3$. For ~bi-Verma (= 2-Verma) modules we give the
general explicit formula for a class of singular vectors, which
exhausts all possible cases for $\cg = sl(2)$.

In Section 4 we first recall the procedure of \Dob\ for the
construction of linear \idos. Then we generalize this procedure
for the construction of ~multilinear \idos. This is a general
result which produces a multilinear \ido\ for every singular
vector of a ~$k$ - Verma module, the procedure of \Dob\ being
obtained for $k=1$.

In Section 5 we study bilinear operators for $G = SL(n,\bbr)$
mentioning also which which results are extendable to
$SL(N,\bbc)$. We give explicit formulae for all bilinear \idos\
for $\cg = sl(2,\bbr)$ and $SL(2,\bbr)$, noting the difference
between the algebra and group invariants. We study in some detail
partial cases, in particular, an infinite hierarchy of even order
\idos\ producing $\han$-differentials for all $n\in 2\bbn$, the
ordinary Schwarzian being the case $n=4$. We also give many
examples for $SL(2,\bbr)$.

In Section 6 we give some examples which illustrate additional
new features of the multilinear \idos\ for ~$k>2\,$.

In the Appendix we have summarized the notions of tensor,
symmetric and universal enveloping algebras.

The end of formulations of: Definiton, Remark, Corollary, quoted
statement without proof is marked with ~\dia~, ~the end of a
Proof is marked with ~\bu~.

\vskip 5mm

\newsec{\quad k - Verma modules}

\newsubsec{} 
Let ~$F ~=\bbc$~ or ~$F ~=\bbr$. Let ~$\cg$~ be a semisimple Lie
algebra over ~$F=\bbc$~ or a split real semisimple Lie algebra
over ~$F=\bbr$. Thus ~$\cg$~ has a triangular decomposition:
~~$\cg ~=~ \cg^+ \oplus \ch \oplus \cg^-$, where ~$\ch$~ is a
Cartan subalgebra of $\cg$ (the maximally non-compact Cartan
subalgebra for $F=\bbr$), ~$\cg^+$, resp., ~$\cg^-$, are the
positive, resp., negative root vector spaces of the root system
~$\D ~=~ \D(\cg,\ch)$, corresponding to the decomposition ~$\D
~=~ \D^+ \cup \D^-$~ into positive and negative roots. [For
$F=\bbr$ this decomposition is a partial case of a Bruhat
decomposition.] In particular, one has: ~$\cg^\pm ~=~
\oplus_{\b\in\D^+} ~\cg^\pm_\b\,$. In our cases 
~$\dim \cg^\pm_\b ~=~ 1$, $\forall \b\in\D^+$, and further
~$X^\pm_\b$~ will denote a vector spanning ~$\cg^\pm_\b\,$. ~Let
~$\D_S$~ be the system of simple roots of $\D$. ~ Let ~$\G^+
\in\ch^*$~ denote the set of dominant weights, i.e.,
~$\nu\in\G^+$~ iff ~$(\nu,\a^\vee_i) \in\bbz_+$ for all
$\a_i\in\D_S\,$. ~Let ~$U(\cg)$~ be the universal enveloping
algebra of ~$\cg$~ with unit vector denoted by ~$1_u\,$. (The
notions of tensor, symmetric and universal enveloping algebras
are recalled in Appendix A.)

Let us recall that a ~{\it Verma module} ~$V^\L$~ is defined as
the HWM over ~$\cg$~ with highest weight ~$\L \in \ch^*$~ and
highest weight vector ~$v_0 \in V^\L$, induced from the
one-dimensional representation ~$V_0 \cong \bbc v_0$~ of
~$U(\cb)$~, where ~$\cb = \cb^+ = \ch \oplus \cg^+$~ is a Borel
subalgebra of ~$\cg$, such that:
\eqn\indb{
\eqalign{ &X ~v_0 ~~=~~ 0 , \quad X\in \cg^+ \cr 
&H ~v_0 ~~=~~ \L(H)~v_0\,, \quad H \in \ch \cr }} Thus one has:
\eqn\verm{ V^\L ~\cong ~ 
U(\cg)\otimes_{U(\cb)}v_0 ~\cong~ U(\cg^-)\otimes v_0 }
(isomorphisms between vector spaces). One considers ~$V^\L$~ as a
left ~$U(\cg)$--module (w.r.t. multiplication in $U(\cg)$).

Verma modules are generically irreducible. A Verma module is
reducible iff it has singular vectors (one or more) \BGG. A
~{\it singular vector}~ of a Verma module ~$V^\L$~ is a vector
~$v_s \in V^\L$, such that ~$v_s ~\notin ~F~1_u \otimes v_0$~
and~:
\eqn\sing{
\eqalign{& X ~v_s ~~=~~ 0 , \quad X\in \cg^+ \cr 
&H ~v_s ~~=~~ (\L(H) -\mu(H))~v_s\,, \quad
\mu \in \G^+\,, ~\mu\neq 0, ~~
H \in \ch \cr }} The space $U(\cg^-)\otimes v_s$~ is a submodule
of ~$V^\L$~ isomorphic to the Verma module ~$V^{\L - \mu} =
U(\cg^-)\otimes v'_0$~ where ~$v'_0$~ is the highest weight
vector of ~$V^{\L - \mu}$; the isomorphism being realized by
~$v_s\mapsto 1_u\otimes v'_0$. Furthermore, there exists (at
least one) decomposition ~$\mu ~=~
\sum_{i=1}^n m_i\b_i\,$, $m_i\in\bbn$, $\b_i\in\D^+$; the latter 
statement in the case ~$n=1$~ means that ~$\mu ~=~ m\b$,
$m\in\bbn$, $\b\in\D^+$. For each such decomposition there
exists a composition of embeddings of the Verma modules ~$V_i
\equiv V^{\L - m_i\b_i}$~ which thus form a nested sequence of
submodules, so that ~$V_i$~ is a submodule of ~$V_{i-1}\,$,
$i=1,\ldots,n$,~ $V_0 \equiv V^\L$. Each such submodule is
generated by a singular vector of weight ~$m_i\b_i$. The
singular vector of weight ~$m\b$~ is given by \Dob~:
\eqn\sina{ v_s ~~=~~ v_s^{m\b} ~~=~~ {\cal P}^{m\b} 
(X^-_1, \dots, X^-_\ell)\otimes v_0 } where ~${\cal P}^{m\b}$~ is
a homogeneous polynomial in its variables of degrees $mn_i$,
where ~$n_i \in \bbz_+$~ come from $\beta = \sum n_i\a_i$,
~$\a_i$ form the system of simple roots $\D_S$. The polynomial
~${\cal P}^\b_m$~ is unique up to a non-zero multiplicative
constant. From this follows that the singular vector of weight
~$\mu$~ is given by:
\eqn\sinb{ v_s ~~=~~ v_s^{\mu} ~~=~~ {\cal P}^{m_n\b_n} ~\ldots ~
{\cal P}^{m_1\b_1} ~\otimes v_0 }

Finally, we should mention that in this setting the highest
weight satisfies:
\eqn\red{ (\L + \r ~, ~\b_1^\vee) - m_1 ~=~ (\L +
\r)(H_{\b_1}) - m_1 ~=~ 0 ~, }
where ~$\r$~ is half the sum of all positive roots, ~ $\a^\vee
\equiv 2\a/(\a,\a)$ for any $\a\in\D$, ~$(~, ~)$~ is the scalar
product in ~$\ch^*$, ~$H_\a\in\ch$~ corresponds to ~$\a\in\D^+$~
under the isomorphism ~$\ch \cong \ch^*$. As a consequence one
has:
\eqn\reda{ (\L_{i-1} + \r ~, ~\b_i^\vee) ~~=~~ m_i ~, ~~i=1,\ldots,n,
~~\L_0 =\L, ~\L_i = \L_{i-1} - m_i\b_i }

One should note that the condition \red\ is in fact necessary and
sufficient for the reducibility of a Verma module. Thus one may
say equivalently that the Verma module ~$V^\L$~ is reducible iff
there exists a root ~$\b \in\D^+$~ and ~$m\in\bbn$~ so that holds
\BGG~:
\eqn\red{ (\L + \r ~, ~\b^\vee) - m ~=~ (\L +
\r)(H_\b) - m ~=~ 0 }
We have chosen a different exposition here since in the
generalization of the Verma modules we introduce below we do not
rely on an analogue of \red\ and reducibility is discussed via
singular vectors.

\newsubsec{} 
We introduce now a generalization of the Verma modules. Let
~$k$~ be a natural number, let ~$\ct_k(\cg)$~ be the tensor
product:
\eqn\kte{ \ct_k(\cg) ~~\doteq ~~T_k(U(\cg)) ~~=~~
\underbrace{ U(\cg) \otimes \dots \otimes U(\cg) }_k }
and let ~$\cs_k(\cg)$~ be the symmetric tensor product:
\eqn\kte{ \cs_k(\cg) ~~\doteq ~~S_k(U(\cg)) ~~=~~
T_k(U(\cg))/I_k(U(\cg)) } Then arbitrary elements of
~$\cs_k(\cg)$~ shall be denoted as follows:
\eqn\elme{
u ~~=~~ \left\{ u_1 \otimes \dots
\otimes u_k \right\} ~, 
\quad u_j \in U(\cg) \,, } 
where ~$\{ ~...~\}$~ denotes the symmetric tensor product which
is preserved under arbitrary permutations ~$u_i ~\llra ~u_j\,$.

\st
{\bf Definition:}~~ {\it A} ~{\bf k - Verma module}~ $_kV^\L$~
{\it is a highest weight module over ~$\cg$~ induced from the
one-dimensional representation of ~$\cb$~ (cf. \indb) so that:
\eqn\kverm{ _kV^\L ~~\cong ~~ 
\cs_k(\cg) ~\hat\otimes ~v_0 
~~\cong~~ \cs_k(\cg^-) ~\hat\otimes ~v_0 \,, \quad
\hat\otimes ~\equiv ~\otimes_{U(\cb)} }
(isomorphisms between vector spaces). ~$_kV^\L$~ is considered a
left ~$U(\cg)$--module (w.r.t. multiplication in $U(\cg)$).
Denoting arbitrary elements ~$v$~ of ~$_kV^\L$~ consistently with
\elme~:
\eqn\elem{ 
v ~~=~~ \left\{ u_1 \otimes \dots
\otimes u_k \right\} ~ \ho ~v_0 \,, \quad u_j \in U(\cg^-) } 
we define the action of $U(\cg)$ as follows:}
\eqn\kact{
X~ v ~~=~~ \sum_{j=1}^k ~\left\{ u_1 \otimes \dots \otimes X u_j
\otimes \dots
\otimes u_k \right\} ~ \ho ~v_0 \,, \quad X\in U(\cg) \,. \quad 
\diamondsuit }

\st
{\it Remark 1:}~~ Clearly, ~1 - Verma modules are usual Verma
modules.~~\dia

\st
{\bf Corollary:} (from the above definition) ~~ {\it Let
~$H\in\ch$, ~let ~$u_j$~ be from the PBW basis of $U(\cg^-)$,
~let ~$\mu_j$~ be the (negative) weight of $u_j$, i.e., ~$[H,u_j]
~=~ -\mu_j(H) u_j\,$. Then we have:}
\eqn\hact{ \eqalign{ 
H~ v ~=&~ \sum_{j=1}^k ~\left\{ u_1 \otimes \dots \otimes H u_j
\otimes \dots
\otimes u_k \right\} ~ \ho ~ v_0 ~=\cr
=& ~ \sum_{j=1}^k ~\left\{ u_1 \otimes \dots \otimes u_j \otimes
\dots
\otimes u_k \right\} ~ \ho ~ \left( H - \mu_j(H) \right) ~ 
v_0 ~=\cr =&~ \left( k \L(H) - \sum_{j=1}^k \mu_j(H) \right) ~v
\, . \quad
\diamondsuit }}

\newsubsec{} 
We need some more notation to proceed further.~ Let ~$P(\mu) ~=~
\#$ of ways $\mu\in\G^+$ can be presented as a sum of positive
roots $\b$. (In general, each root should be taken with its
multiplicity; however, in the cases here all multiplicities are
equal to 1.) By convention ~$P(0) ~\equiv~ 1$. ~Let ~$_k\G^+
~\doteq ~ \G^+ \times \dots \times \G^+\,$, $k$ factors. ~Let
~$_k\mu ~=~ (\mu_1,\dots,\mu_k) \in {_k\G}^+\,$, $\mu_j\in
\G^+\,$, and let ~$\s(_k\mu) ~\doteq ~\sum_{j=1}^k \mu_j ~
\in\G^+\,$. ~Let ~$\mu\in\G^+$~ and let ~ 
$P_k(\mu) ~=~ \#$ of elements ~$_k\nu ~=~ (\nu_1,\dots,\nu_k)
~\in {_k\G}^+\,$, such that ~$\s(_k\nu) ~=~ \mu$, each such
element being taken with multiplicity ~$\prod_{j=1}^k ~P(\nu_j)$.
~Clearly, ~$P_k(0) ~=~ 1$, ~$P_k(\b) ~=~ k$, $\forall \b\in\D_S$.
Finally we define:
\eqn\aaa{ _k\G^+_>~~\doteq ~~ \{ 
_k\mu = (\mu_1,\dots,\mu_k) \in {_k\G}^+ ~\vert ~
\mu_1 \geq \ldots \geq \mu_k \} }
where some ordering of ~$\ch^*\,$, (e.g., the lexicographical
one), is implemented. ~Let ~$\mu\in\G^+$~ and let ~ $P^>_k(\mu)
~=~ \#$ of elements ~$_k\nu ~=~ (\nu_1,\dots,\nu_k) ~\in
{_k\G}^+_>\,$, such that ~$\s(_k\nu) ~=~ \mu$, each such element
being taken with multiplicity ~$\prod_{j=1}^k ~P(\nu_j)$.
~Clearly, ~$P_k(0) ~=~ 1$, ~$P^>_k(\b) ~=~ 1$, $\forall
\b\in\D_S$. Now one can prove the following:

\st
{\bf Proposition 1:}~~ {\it Let $\L\in\ch^*$ and let
\eqn\lev{ 
_kV^\L_\mu ~~\doteq ~~ \{ ~v ~\in~ {_kV}^\L ~\vert ~ H~v ~=~
(k\L(H) - \mu(H))~v ~\} \,.} Then we have:
\eqna\deco
$$\eqalignno{ & _kV^\L ~~=~~ \mathop{\oplus}\limits_{\mu\in\G^+}
~ _kV^\L_{\mu} &\deco a\cr &&\cr &\dim ~{_kV}^\L_{\mu} ~~=~~
P^>_k(\mu) &\deco b\cr &&\cr &&\cr &\eqalign{ _kV^\L_{\mu} ~~=&~~
\sum_{ _k\nu ~\in ~{_k\G}^+_>\atop { \atop \s(_k\nu) ~=~ \mu }}
~\sum_{{{\b^1_i \in \D^+ \atop
\Sigma\b^1_i = \nu_1}\atop \beta^1_1\leq \beta^1_2\leq \dots \leq
\beta^1_{n_1}}}
~\dots ~\sum_{{{\b^k_i \in \D^+ \atop
\Sigma\b^k_i = \nu_k}\atop \beta^k_1\leq \beta^k_2\leq \dots \leq
\beta^k_{n_k}}} ~\times \cr &\times ~ \left\{ 
X^-_{\beta^1_1} \dots X^-_{\beta^1_{n_1}} ~\otimes ~
\dots ~\otimes ~ 
X^-_{\beta^k_1} \dots X^-_{\beta^k_{n_k}} \right\} ~ \ho ~ F ~
v_0
\cr} &\deco c\cr &&\cr &&\cr 
& _kV^\L_0 ~~=~~
\left\{ 1_u \otimes \dots \otimes 1_u \right\} ~\ho ~F ~v_0
&\deco d\cr &&\cr &_kV^\L ~~=~~ \cs_k(\cg^-)~ {_kV}^\L_0 \,,
\qquad
\cg^+\, {_kV}^\L_0 ~=~ 0 
&\deco e\cr }$$ where in \deco{c} the ordering of the root system
inherited from the ordering of ~$\ch^*$~ is implemented.} \nl
{\it Proof:}~~ Completely analogous to the classical case $k=1$
\Di.~\bu

\newsec{\quad Singular vectors of\ k - Verma modules}

\newsubsec{} 
In contrast to the ordinary Verma modules ($k=1$), the $k$ -
Verma modules for ~$k\geq 2$ are reducible independently of the
highest weight, which is natural taking into account their tensor
product character. This we show by exhibiting singular vectors
for arbitrary highest weights.

We call a ~{\it singular vector}~ of a $k$ - Verma module
~$_kV^\L$~ a vector ~$v_s \in~ {_kV}^\L$, such that ~$v_s
\notin {_kV}^\L_0$~ and~:
\eqna\sinc
$$\eqalignno{& X ~v_s ~~=~~ 0 , \quad X\in \cg^+ &\sinc a\cr &H
~v_s ~~=~~ (k\L(H) -\mu(H))~v_s\,, \quad
\mu \in \G^+\,, ~\mu\neq 0, ~~
H \in \ch &\sinc b\cr }$$ i.e., ~$v_s$~ is homogeneous: ~$v_s
\in~{_kV}^\L_\mu$~ for some $\mu\in\G^+$. For $k=1$ \sinc{}
coincide with \sing.

The space ~$\cs_k(\cg^-)~v_s$~ is a submodule of ~$_kV^\L$~
isomorphic to the Verma module ~$_kV^{k\L - \mu} ~=~
\cs_k(\cg^-)\otimes v'_0$~ where ~$v'_0$~ is the highest weight
vector of ~$_kV^{k\L - \mu}$; the isomorphism being realized by
~$v_s \mapsto \{ 1_u \otimes
\ldots 1_u \} \ho v'_0$. 

In the next two Subsections we show some explicit examples for
the cases ~$k=2,3$.

\newsubsec{} 
We consider now the case ~$k=2$~ i.e., ~{\it bi-Verma} (=
2-Verma) modules. We take a weight ~$\mu = n\a\,$, where
~$n\in\bbn$~ and ~$\a\in\D_S$~ is any simple root. We have
~$\dim~ {_2V}^\L_{n\a} ~=~ [n/2] +1$, where $[x]$ is the biggest
integer not exceeding $x$. The possible singular vectors have the
following form:
\eqn\sind{
_2v_s^{n\a} ~~=~~ \sum_{j=0}^{[n/2]} ~
\g^\L_{nj}~ \left\{ \left( X^-_\a \right)^{n-j}
~\otimes ~\left( X^-_\a \right)^{j}
\right\} ~\ho ~v_0 }
The coefficients ~$\g^\L_{nj}$~ are determined from the condition
\sinc{a} with ~$X = X^+_\a$~ $-$ all other cases of \sinc{} are
fulfilled automatically. Thus we have:

\st {\bf Proposition 2:} ~~
{\it The singular vectors of the ~ bi-Verma (= 2-Verma) module
~$_2V^\L$~ of weight ~$\mu = n\a\,$, where ~$n\in\bbn$~ and
~$\a\in\D_S$~ is any simple root, are given by formula \sind\
with the coefficients ~$\g_{nj}$~ given explicitly 
(up to multiplicative renormalization) by:
\eqn\verc{\eqalign{ 
&\g^\L_{nj} ~~=~~ \g_0 ~\g(n,\L(H))~ (-1)^j ~(2-\d_{j,n/2}) ~ {n
\choose j} ~ {\G(\L(H) +1-n+j) ~\G(\L(H) +1-j) \over
\G(\L(H) +1-n) ~\G(\L(H) +1) } \cr &\cr 
&\g(n,\L(H)) ~~=~~ \cases{ 1, & ~for ~$n$~ even and arbitrary
~$\L(H)$ \cr 1, & ~for ~$n$~ odd, ~$\L(H) ~=~
n-1,n-2,\ldots,(n-1)/2 $ \cr 0, & ~for ~$n$~ odd, ~$\L(H) ~\neq
~n-1,n-2,\ldots,(n-1)/2 $ \cr}
\cr} }
and $\g_0$ is an arbitrary non-zero constant.} \nl {\it Proof:}~~
By direct verification.~ \bu

We give the lowest cases of the above general formula for
illustration (fixing the overall constant $\g_0$ appropriately)~;
\eqna\sngt
$$\eqalignno{ &_2v_s^{\a} ~~=~~ \left\{ X^-_\a ~\otimes 1_u
~\right\} ~\ho ~v_0
\,, \quad \L(H) ~=~ 0
&\sngt a\cr &_2v_s^{2\a} ~~=~~ \left\{\L(H) ~ \left( X^-_\a
\right)^2 ~\otimes ~ 1_u ~-~ (\L(H)-1) ~ X^-_\a ~\otimes ~ X^-_\a
\right\} ~\ho ~v_0 \,, \quad \forall \L(H) ~~~~ &\sngt b\cr
&\eqalign{ _2v_s^{3\a} ~~=&~~ \big\{\L(H) ~ \left( X^-_\a
\right)^3 ~\otimes ~ 1_u ~-\cr &~- 3 ~(\L(H)-2) ~ \left( X^-_\a
\right)^2 ~\otimes ~ X^-_\a \big\} ~\ho ~v_0 \,, \qquad \L(H) ~=~
1,2 \cr} &\sngt c\cr &\eqalign{ _2v_s^{4\a} ~~=&~~ \Big\{
\L(H)~(\L(H)-1) ~ \left( X^-_\a \right)^4 ~\otimes ~ 1_u ~-\cr
&-~ 4 ~(\L(H)-1) ~(\L(H)-3) ~
\left( X^-_\a \right)^3
~\otimes ~ X^-_\a ~+\cr &+~ 3 ~(\L(H)-2) ~(\L(H)-3) ~
\left( X^-_\a \right)^2
~\otimes ~ \left( X^-_\a \right)^2 \Big\} ~\ho ~v_0 \,, \qquad
\forall \L(H) \cr} &\sngt d\cr }$$

Proposition 2 confirms that ~bi-Verma~ modules are always
reducible since they possess singular vectors independently of
$\L$. In fact, they have an infinite number of singular vectors
of weights ~$n\a_i$~, for any even positive integer ~$n$~ and any
simple root ~$\a_i$~. ~Noreover, they possess singular vectors of
other weights, also independent of $\L$. For example we consider
weights ~$\mu_n ~=~n \b ~=~n(\a_1 + \a_2)$, where ~$\b$~ is a
positive root, and ~$\a_1\,,\,,\a_2\,,$~ are two simple roots,
e.g., of equal minimal length (for simplicity). Then there exist
singular vectors of these weights given by, e.g.:
\eqn\sngss{
\eqalign{ 
_2v_s^{\b} ~~=&~~ \Big\{ \L_1 ~ X^-_1 ~ X^-_2 ~\otimes ~ 1_u ~-~
\L_2~ ~ X^-_2 ~ X^-_1 ~\otimes ~ 1_u ~-\cr &-~
\left( \L_1 + \L_2 +1 \right) ~
X^-_1 ~\otimes ~ X^-_2
\Big\} ~\ho ~v_0 ~, \qquad \forall \L \cr
\L_a ~&\equiv~ \L(H_a) ~, ~~ a=1,2 
} }
\eqna\sbi 
$$\eqalignno{ _2v_s^{2\b} ~~=&~~ \Big\{ a_1 ~(X^-_3)^2 ~\otimes ~1_u ~+~
a_2 ~X^-_2 X^-_3 X^-_1 ~\otimes ~1_u ~+~ a_3 ~(X^-_2)^2 (X^-_1)^2
~\otimes ~1_u ~+ &\cr &~~+ b_1 ~X^-_3 X^-_1 ~\otimes ~X^-_2 ~+~
b_2 ~X^-_2 X^-_3 ~\otimes ~X^-_1 ~+ &\cr &~~+ c_1 ~X^-_2
(X^-_1)^2 ~\otimes ~X^-_2 ~+~ c_2 ~(X^-_2)^2 X^-_1 ~\otimes
~X^-_1 ~+ &\cr &~~+ d_1 ~X^-_3 ~\otimes ~X^-_3 ~+~ d_2 ~X^-_3
~\otimes ~X^-_2 X^-_1 ~+ &\cr &~~+ d_3 ~X^-_2 X^-_1 ~\otimes
~X^-_2 X^-_1 ~+~ d_4 ~(X^-_2)^2 ~\otimes ~(X^-_1)^2 \Big\} ~\ho ~v_0
~, \qquad
\forall \L &\sbi a\cr } $$ where for the two solutions (present
in this case) the coefficients are: $$\eqalignno{ a_1 ~~=&~~
\L_1^2 \L_2 (\L_1 + \L_2 + 1) (\L_2 + 1) 
&\cr a_2 ~~=&~~ - \L_1 \L_2 (\L_1 + \L_2 + 1) (\L_1 - \L_2 - 2)
&\cr a_3 ~~=&~~ - \L_1 \L_2 (\L_1 + \L_2 + 1) &\cr b_1 ~~=&~~ -
\L_1 (\L_1 + \L_2 + 1) (\L_1 + \L_2) (\L_2 + 2) &\cr b_2 ~~=&~~
\L_1 \L_2 (\L_1 + \L_2 + 1) (\L_1 + \L_2) 
&\cr c_1 ~~=&~~ \L_1 (\L_1+\L_2+1) (\L_1+\L_2) &\cr c_2 ~~=&~~
\L_2 (\L_1 + \L_2 + 1) (\L_1 + \L_2) 
&\cr d_1 ~~=&~~ - \L_1^2 (\L_1 + \L_2) (\L_2^2 + \L_2 + 1) &\cr
d_2 ~~=&~~
\L_1 (\L_1 + \L_2) (\L_1 \L_2 + 2 \L_1 - \L_2^2 + \L_2) 
&\cr d_3 ~~=&~~ - (\L_1 + \L_2) (\L_1^2 + \L_1 \L_2 + \L_2^2)
&\cr d_4 ~~=&~~ 0 &\sbi b\cr }$$ $$\eqalignno{ a_1 ~~=&~~ 2
\L_1^2 \L_2 (\L_1 + \L_2 + 1) (\L_2^2 - \L_2 - 1 + \L_1 \L_2)
&\cr a_2 ~~=&~~ - 2 \L_1 (\L_1 + \L_2 + 1) (\L_1 + \L_2 - 1)
(\L_1 - \L_2) (\L_2 + 1) &\cr a_3 ~~=&~~
\L_1 (\L_1 + \L_2 + 1) (\L_1 + \L_2 - 1) (\L_1 - \L_2) 
&\cr b_1 ~~=&~~ - 2 \L_1 (\L_1 + \L_2 + 1)^2 (\L_1 + \L_2) (\L_2
- 1) &\cr b_2 ~~=&~~ 2 \L_1 (\L_1 + \L_2 + 1) (\L_1 + \L_2)
(\L_2^2 - 2 \L_2 - 1 +
\L_1 \L_2 + \L_1) 
&\cr c_1 ~~=&~~ 0 &\cr c_2 ~~=&~~ - 2 (\L_1 + \L_2 + 1) (\L_1 +
\L_2 - 1) (\L_1 + \L_2) (\L_1 - \L_2) &\cr d_1 ~~=&~~ - 2 \L_1^2
(\L_1 + \L_2) (\L_2 - 1) (\L_1 \L_2 + \L_1 + \L_2^2) &\cr d_2
~~=&~~ - 2 \L_1 (\L_1 + \L_2) (\L_2 - 1) (\L_2^2 - 2 \L_2 - 1 -
\L_1^2) &\cr d_3 ~~=&~~ - 2 (\L_1 + \L_2) (\L_2 - 1) (\L_1 \L_2 +
\L_1 + \L_2^2) &\cr d_4 ~~=&~~ \L_1 (\L_1+\L_2) (\L_1+\L_2+1)^2
&\sbi c\cr }$$

Furthermore, the structure of the Verma modules is additionally
complicated since there is no factorization of the singular
vectors as for $k=1$, cf. \sinb. For the latter consider the
weight ~$\mu_0 ~=~ 2\a_1 ~+~ \a_2\,$, ~with ~$\a_1\,,\,\a_2$~ as
in the previous example. Then there exists a singular vector of
this weight given by:
\eqn\sngsr{
\eqalign{ 
_2v_s^{\mu_0} ~~=&~~ \Big\{ \left( 2\L_1 \L_2 + \L_1 - \L_2 -1
\right) ~ \left( X^-_1 \right)^2 ~ X^-_2 ~\otimes ~ 1_u ~-~ 2
\L_1 \L_2 ~ X^-_1 ~ X^-_2 ~ X^-_1 ~
\otimes ~ 1_u ~+\cr &+~ 
\left( \L_1 + \L_2 +1 \right) ~
\left( X^-_1 \right)^2 ~\otimes ~ X^-_2
~-~ 2 \left( \L_1 -1 \right) ~\left( \L_2 +1 \right) ~ X^-_1 ~
X^-_2 ~ \otimes ~X^-_1 ~+ \cr &+~ 2 \L_2 \left( \L_1 -1 \right) ~
X^-_2 ~ X^-_1 ~ \otimes ~X^-_1
\Big\} ~\ho ~v_0 \,, \qquad \forall ~\L 
\cr} }

\newsubsec{}
Here we consider the case $k=3$. i.e., ~{\it tri-Verma} (=
3-Verma) modules over arbitrary ~$\cg$. Consider a weight ~$\mu
= n\a\,$, where ~$n\in\bbn$~ and ~$\a\in\D_S$~ is any simple
root. We first note the dimension of the weight space:
\eqn\dimt{ 
\dim~ {_3V}^\L_{n\a} ~~=~~ \left(n -3 \left[\has\right]\right) ~
\left(1 + \left[\has\right] \right) 
~+~ \d_{\has,\left[\has\right]} } The possible singular vectors
have the following form:
\eqn\sind{
_3v_s^{n\a} ~~=~~ \sum_{{ j,k \in\bbz_+ \atop n-j-k \geq j \geq
k} }
\g^\L_{njk}~ \left\{ \left( X^-_\a \right)^{n-j-k}
~\otimes ~\left( X^-_\a \right)^{j} ~\otimes ~\left( X^-_\a
\right)^{k}
\right\} ~\ho ~v_0 } 
The coefficients ~$\g^\L_{njk}$~ are determined form the
condition \sinc{a} with ~$X = X^+_\a$~ - all other cases in of
\sinc{} are fulfilled automatically. We give now the singular
vectors for ~$n\leq 6$~ denoting ~$\hL ~\equiv~\L(H)$~:
\eqna\sngr
$$\eqalignno{ &_3v_s^{\a} ~~=~~ \big\{ X^-_\a ~\otimes ~ 1_u
~\otimes ~ 1_u
\big\} ~\ho ~v_0
\,, \qquad \hL ~=~ 0 &\sngr a\cr}$$
$$\eqalignno{ &\eqalign{ _3v_s^{2\a} ~~=~~ &\big\{\hL ~ \left(
X^-_\a \right)^2 ~\otimes ~ 1_u ~\otimes ~ 1_u ~-\cr &-~ (\hL-1)
~ X^-_\a ~\otimes ~ X^-_\a ~\otimes ~ 1_u \big\} ~\ho ~v_0 \,,
\qquad \forall \hL \cr} &\sngr b\cr}$$ $$\eqalignno{ &\eqalign{
_3v_s^{3\a} ~~=~~ &\Big\{\hL^2 ~ \left( X^-_\a \right)^3 ~\otimes
~ 1_u ~\otimes ~ 1_u ~-\cr &-~ 3 ~\hL ~(\hL-2) ~
\left( X^-_\a \right)^2
~\otimes ~ X^-_\a ~\otimes ~ 1_u ~+\cr &+~ 2 ~(\hL-1) ~(\hL-2) ~
X^-_\a ~\otimes ~ X^-_\a ~\otimes ~ X^-_\a
\Big\} 
~\ho ~v_0 \,, \qquad \forall \hL \cr} &\sngr c\cr}$$
$$\eqalignno{ &\eqalign{ _3v_s^{4\a} ~~=~~ &\Big\{ \hL~(\hL-1) ~
\left( X^-_\a \right)^4 ~\otimes ~ 1_u ~\otimes ~ 1_u ~-\cr &-~ 4
~(\hL-1) ~(\hL-3) ~
\left( X^-_\a \right)^3
~\otimes ~ X^-_\a ~\otimes ~ 1_u ~+\cr &+~ 3 ~(\hL-2) ~(\hL-3) ~
\left( X^-_\a \right)^2
~\otimes ~ \left( X^-_\a \right)^2 ~\otimes ~ 1_u \Big\} ~\ho
~v_0 \,, \qquad \forall \hL \cr} &\sngr d\cr}$$ $$\eqalignno{
&\eqalign{ _3v'^{4\a}_s ~~=~~ &\Big\{ ~ \left( X^-_\a \right)^4
~\otimes ~ 1_u ~\otimes ~ 1_u ~+~ 8 ~\left( X^-_\a \right)^3
~\otimes ~ X^-_\a ~\otimes ~ 1_u ~+\cr &+~ 12 ~ \left( X^-_\a
\right)^2 ~\otimes ~ X^-_\a ~\otimes ~ X^-_\a
\Big\} 
~\ho ~v_0 \,, \qquad \hL ~=~ 1 \cr} &\sngr {d'}\cr}$$
$$\eqalignno{ &\eqalign{ _3v_s^{5\a} ~~=~~ &\Big\{ \hL^2~(\hL-1)
~ \left( X^-_\a \right)^5 ~\otimes ~ 1_u ~\otimes ~ 1_u ~-\cr &-~
5 ~\hL ~(\hL-1) ~(\hL-4) ~
\left( X^-_\a \right)^4
~\otimes ~ X^-_\a ~\otimes ~ 1_u ~+\cr &+~ 2 ~\hL ~(\hL-3)
~(\hL-4) ~
\left( X^-_\a \right)^3
~\otimes ~ \left( X^-_\a \right)^2 ~\otimes ~ 1_u ~+\cr &+~ 8
~(\hL-1) ~(\hL-3) ~(\hL-4) ~
\left( X^-_\a \right)^3
~\otimes ~ X^-_\a ~\otimes ~ X^-_\a ~-\cr &-~ 6 ~(\hL-2) ~(\hL-3)
~(\hL-4) ~
\left( X^-_\a \right)^2
~\otimes ~ \left( X^-_\a \right)^2 ~\otimes ~ X^-_\a \Big\} ~\ho
~v_0 \,, \quad \forall \hL \cr} &\sngr e\cr}$$ $$\eqalignno{
&\eqalign{ _3v'^{5\a}_s ~~=~~ &\Big\{ ~2~ \left( X^-_\a \right)^3
~\otimes ~ \left( X^-_\a \right)^2 u ~\otimes ~ 1_u ~-\cr &-~
\left( X^-_\a \right)^3 ~\otimes ~ X^-_\a ~ \otimes ~X^-_\a ~
\Big\} 
~\ho ~v_0 \,, \qquad \hL ~=~ 2 \cr} &\sngr {e'}\cr}$$
$$\eqalignno{ &\eqalign{ _3v_s^{6\a} ~~=~~ &\Big\{ \hL ~(\hL-1)
~(\hL-2) ~ \left( X^-_\a \right)^4 ~\otimes ~ ~ \left( X^-_\a
\right)^2 ~\otimes ~ 1_u ~-\cr &-~ (\hL-1)^2 ~(\hL-2) ~
\left( X^-_\a \right)^4
~\otimes ~ X^-_\a ~
\otimes ~ X^-_\a ~-\cr 
&-~ \hL ~(\hL-1) ~(\hL-3) ~
\left( X^-_\a \right)^3
~\otimes ~ \left( X^-_\a \right)^3 ~\otimes ~ 1_u ~+\cr &+~ 2
~(\hL-1) ~(\hL-2) ~(\hL-3) ~
\left( X^-_\a \right)^3 
~\otimes ~ \left( X^-_\a \right)^2 ~\otimes ~ X^-_\a ~-\cr &-~
(\hL-2)^2 ~(\hL-3) ~
\left( X^-_\a \right)^2
~\otimes ~ \left( X^-_\a \right)^2 ~\otimes ~
\left( X^-_\a \right)^2 \Big\} 
~\ho ~v_0 \,, \qquad \forall \hL \cr} &\sngr f\cr }$$
$$\eqalignno{ &\eqalign{ _3v'^{6\a}_s ~~=~~ &\Big\{ \hL ~(\hL-1)
~(\hL-2)
\left( X^-_\a \right)^6
~\otimes ~ 1_u ~ \otimes ~ 1_u ~-\cr &-~ 6 ~ (\hL-1) ~ (\hL-2)
~(\hL-5) ~ \left( X^-_\a \right)^5 ~\otimes ~ X^-_\a ~\otimes ~
1_u ~+\cr &+~ 15~ (\hL-2) ~ (\hL-4) ~(\hL-5) ~ \left( X^-_\a
\right)^4 ~\otimes ~ \left( X^-_\a \right)^2 ~\otimes ~ 1_u ~-\cr
&-~ 10 ~(\hL-3) ~(\hL-4) ~(\hL-5) ~
\left( X^-_\a \right)^3 
~\otimes ~ \left( X^-_\a \right)^3 ~\otimes ~ 1_u \Big\} ~\ho
~v_0 \,, \quad \forall \hL \cr} &\sngr {f'}\cr }$$

We give those examples in order to point out some new features
appearing for ~tri-Verma modules in comparison with the ~bi-Verma
modules: \item{\bu} {\it Independently of ~$\L(H)$~ there exists
a singular vector at ~{\it any level}~ $n\a$, except the lowest
$n=1$, while for ~bi-Verma modules singular vectors at ~{\it
odd}~ levels exist only for special values of ~$\L(H)\,$;
\item{\bu} 
There exists ~{\it more than one singular vector}~ at any fixed
level ~$n\a$~ for ~$n \geq 6$~ and arbitrary ~$\L(H)$. For
special values of ~$\L(H)$~ there exists a second singular vector
for ~$n=4,5\,$. }

Similar facts hold for ~$k$ - Verma modules for $k>3$. These
questions will be considered in another publication. In the
present paper we would like to demonstrate on examples the
usefulness of these modules, which we do in the next Sections.

\vskip 5mm

\newsec{\quad Multilinear \idos} 

\newsubsec{} 
We start here by sketching the procedure of \Dob\ for
construction of ~{\it linear} \idos\ which we generalize in the
next subsection for ~{\it multilinear} \idos. Let ~$G$~ be a
semisimple Lie group and let ~$\cg$~ denote its Lie algebra.
(Note that the procedure works in the same way for a reductive
Lie group, since only its semisimple subgroup is essential for
the construction of the \idos. We restrict to semisimple groups
for simplicity. For more technical simplicity one may assume that
in addition ~$G$~ is linear and connected.) Let ~$G ~=~ KA_0N_0$~
be an Iwasawa decomposition of $G$, where ~$K$~ is the maximal
compact subgroup of $G$, ~$A_0$~ is abelian simply connected, the
so-called vector subgroup of ~$G$, ~$N_0$~ is a nilpotent simply
connected subgroup of ~$G$~ preserved by the action of ~$A_0$.
Further, let $M_0$ be the centralizer of $A_0$ in $K$. ($M_0$
has the structure $M_0 = M_0^d M_0^r$, where $M_0^d$ is a finite
group, $M_0^r$ is reductive with the same Lie algebra as $M_0$.)
Then ~$P_0 = M_0A_0N_0$~ is called a ~{\it minimal parabolic
subgroup}~ of $G$. A ~{\it parabolic subgroup}~ of ~$G$~ is any
subgroup which is isomorphic to a subgroup ~$P ~=~ MAN$~ such
that: ~$P_0 \subset P \subset G$, ~$M_0 \subset M$, ~$A_0 \supset
A$, ~$N_0 \supset N$. ~ [Note that in the above considerations
every subgroup ~$N$~ may be exchanged with its Cartan conjugate
~$\tn$.] ~ The number of non-conjugate parabolic subgroups
(counting also the case ~$P=G=M$) is ~$2^{r_0}\,$, ~$r_0 ~=~ \dim
A_0\,$.

Parabolic subgroups are important because the representations
induced from them generate all admissible irreducible
representations of $G$ \La, \KZ. In fact, for this it is enough
to use only the so-called ~{\it cuspidal}~ parabolic subgroups,
singled out by the condition that ~rank$\, M =$ rank$\, M\cap K$;
thus $M$ has discrete series representations.

Let ~$P$~ be a cuspidal parabolic subgroup and let ~$\mu$~ fix a
discrete series representation ~$D^\mu$~ of ~$M$~ on the Hilbert
space ~$V_\mu$~ or the so-called limit of a discrete series
representation (cf. \Kn). Let ~$\nu$~ be a (non-unitary)
character of $A$, $\nu\in\ca^*$, where ~$\ca$~ is the Lie algebra
of $A$. We call the induced representation $\chi =$
Ind$^G_P(\mu\otimes\nu \otimes 1)$ an ~{\it elementary
representation} of $G$. (These are called {\it generalized
principal series representations} (or limits thereof) in \Kn.)
Consider now the space of functions
\eqn\fun{
\cc_\chi ~~=~~ \{ \cf \in C^\infty(G,V_\mu) ~\vr ~ 
\cf (gman) ~=~ e^{\nu(H)} D^\mu(m^{-1}) \cf (g) \} } 
where ~$g\in G$, $m\in M$, $a= \exp(H)$, $H\in\ca$, $n\in N$.
~The special property of the functions of $\cih$ is called ~{\it
right covariance} \Dob\ (or {\it equivariance}). It is well known
that ~$\cc_\chi$~ can be thought of as the space of smooth
sections of the homogeneous vector bundle (called also vector
~$G$-bundle) with base space ~$G/P$~ and fibre ~$V_\mu\,$, (which
is an associated bundle to the principal ~$P$-bundle with total
space ~$G$).

Then the elementary representation (ER) ~$\ct^\chi$~ acts in
~$\cc_\chi$, as the left regular representation (LRR), by:
\eqn\lrr{ (\ct^\chi(g)\cf) (g') ~=~ \cf (g^{-1}g') ~, \quad
g,g'\in G } (In practice, the same induction is used with
non-discrete series representations of $M$ and also with
non-cuspidal parabolic subgroups.) One can introduce in
~$\cc_\chi$~ a Fr\'echet space topology or complete it to a
Hilbert space (cf. \Kn). The ERs differ from the LRR (which is
highly reducible) by the specific representation spaces
$\cc_\chi$. In contrast, the ERs are generically irreducible. The
reducible ERs form a measure zero set in the space of the
representation parameters $\mu$, $\nu$. (Reducibility here is
topological in the sense that there exist nontrivial (closed)
invariant subspace.) Next we note that in order to obtain the
\idos\ one may consider the infinitesimal version of \lrr,
namely:
\eqn\lrri{ (\cl^\chi(X)~\cf) (g) ~~\doteq~~ {d\over dt} \cf
(\exp(-tX)g)\vr_{t=0} ~, \quad X\in\cg }

The feature of the ERs which makes them important for our
considerations in \Dob\ and here is a highest weight module
structure associated with them. [It would be lowest weight
module structure, if one exchanges ~$N$~ with ~$\tn$, as is
actually done in \Dob.] Let ~$\gc=\cg$~ if $\cg$ is complex or
real split, otherwise let ~$\gc$~ be the complexification of
$\cg$ with triangular decomposition: ~$\gc = \gc^+ \oplus \ch_c
\oplus \gc^-$~. ~ Further introduce the right action of ~$\gc$~
by the standard formula:
\eqn\rac{ ({\hat X}\cf) (g) ~~=~~ {d\over dt} \cf
(g\exp(tX))\vr_{t=0} } 
where, $X\in \gc$, ~$\cf\in \cc_\chi$, ~$g\in G$, which is
defined first for $X\in\cg$ and then is extended to ~$\gc$~ by
linearity. Note that this action takes ~$\cf$~ out of ~$\cih$~
for some $X$ but that is exactly why it is used for the
construction of the \idos.

We illustrate the highest weight module structure in the case of
the minimal parabolic subgroup. In that case $M=M_0$ is compact
and $V_\mu$ is finite dimensional. Consider first the case when
$M_0$ is non-abelian. Let ~$v_0$~ be the highest weight vector
of $V_\mu\,$. Now we can introduce ~$\bbc$-valued realization
~$\tcih$~ of the space $\cih$ by the formula:
\eqn\sca{ \vf (g) ~~\equiv ~~ \lg ~v_0 ~, ~\cf(g) ~\rg } 
where ~$\lg , \rg$~ is the $M_0$-invariant scalar product in
$V_\mu$. On these functions the counterpart ~$\tih$~ of the LRR
\lrr, its infinitesimal form \lrri\ and the right action of
~$\gc$~ are defined as inherited from the functions $\cf$~:
\eqna\raca
$$\eqalignno{
\tih (g) ~\vf ~~&\doteq ~~ \lg ~v_0 ~, ~{\cal T}^\chi (g) ~ \cf~
 \rg &\raca a\cr \tcl^\chi(X)~\vf ~~&\doteq ~~ \lg ~v_0 ~,
~\cl^\chi(X)~ \cf~ \rg &\raca b\cr {\hat X} ~\vf ~~&\doteq ~~
\lg ~v_0 ~, ~{\hat X} ~\cf ~ \rg &\raca b\cr }$$ In the geometric
language we have replaced the homogeneous vector bundle with base
space ~$G/P$~ and fibre ~$V_\mu\,$~ with a line bundle again with
base space ~$G/P$ (also associated to the principal ~$P$-bundle
with total space ~$G$). The functions ~$\vf$~ can be thought of
as smooth sections of this line bundle. ~ If $M_0$ is abelian or
discrete then $V_\mu$ is one--dimensional and ~$\tcih$~ coincides
with ~$\cih$; so we set ~$\vf ~=~ \cf$. Part of the main result
of \Dob\ is:

\st 
{\bf Proposition:} ~~{\it The functions of the ~$\bbc$-valued
realization ~$\tcih$~ of the ER ~$\cih$~ satisfy~:
\eqna\low 
$$\eqalignno{& {\hat X} \vf ~~=~~ \L(X) ~ \vf ~,
\quad X \in \ch_c &\low a\cr &{\hat X} \vf ~~=~~ 0 ~, \quad X \in
\gc^+ &\low b\cr }$$ where $\L= \L_{\chi} \in (\ch_c)^*$ is built
canonically from $\chi$ and contains all the information from
$\chi$, except about the character $\eps$ of the finite group
$M_0^d$).}~~\dia

Now we note that conditions \low{} are the defining conditions
for the highest weight vector of a highest weight module (HWM)
over $\gc$ with highest weight $\L$, in particular, of a Verma
module with this highest weight, cf. \indb.

Let the signature ~$\chi$~ of an ER be such that the
corresponding ~$\L=\L_\chi$~ satisfies \red\ for some
~$\b\in\D^+$~ and some ~$m\in\bbn$. If $\b$ is a real root, then
some conditions are imposed on the character $\eps$ representing
the finite group $M_0^d$ \SV. Then there exists a \ido\ \Dob~:
\eqna\intd
$$\eqalignno{ &\cd^{m\b} ~~:~~ \tcih ~\lra ~\tcihp &\intd a\cr
&\cd^{m\b} \circ \tih (g) ~~=~~ \tihp (g) \circ \cd^{m\b} ~,
\quad \forall g\in G &\intd a\cr }$$ 
where ~$\chi'$~ is uniquely determined so that ~$\L' ~=~
\L_{\chi'} ~=~ \L - m\b$.

The important fact is that \intd{} is explicitly given by \Dob~:
\eqn\ints{ \cd^{m\b} ~\vf(g) ~~=~~ 
P^{m\b} (\hat X^-_1,\ldots, \hat X^-_\ell) ~\vf (g) } where
~$P^{m\b}$~ is the same polynomial as in \sina\ and $\hat X^-_j$
denotes the action \raca{b}. One important technical
simplification is that the \idos\ \ints\ are ~{\it scalar}
operators since they intertwine two ~{\it line} bundles
~$\tcih\,$, $\tcihp\,$.

\newsubsec{} 
Now we generalize the above sketched construction of \Dob\ to
~{\it multilinear \idos}. We have the following.

\st
{\bf Proposition 3:}~~ {\it Let the signature ~$\chi$~ of an ER
be such that the ~$k$ - Verma module ~$_kV^\L$~ with highest
weight ~$\L=\L_\chi\,$, ~ has a singular vector ~$_kv^\mu_s ~\in
~{_kV}^\L_\mu $, i.e., \sinc{} is satisfied for some
~$\mu\in\G^+$. Let us denote:
\eqn\sinf{ _kv^\mu_s ~~=~~ _k\cp^\mu ~\ho ~v_0 }
where ~$_k\cp^\mu \in \cs_k(\cg^-)$~ is some concrete polynomial
as in \deco{c}. Then there exists a multilinear \ido\ which we
denote by ~$_k\ci^\L_\mu$~ such that:
\eqn\imc{ _k\ci^\L_\mu ~~:~~ \underbrace{ \vf
\otimes \dots \otimes \vf }_k 
 ~ \mt ~ \psi ~, \quad \vf \in \tcih, ~ \psi \in \tcihp }
\eqn\intc{ _k\ci^\L_\mu ~\circ~ \sum_{j=1}^k 
\underbrace{ 1_u \otimes \dots \otimes \tcl^\chi (X) 
\otimes \dots \otimes 1_u }_k 
~~=~~ \tcl^{\chi'} (X) ~\circ ~ {_k\ci}^\L_\mu ~, \quad
\forall X\in \cg }
where ~$\chi'$~ is uniquely determined (up to the representation
parameters of the discrete subgroup $M^d$) so that ~$\L' ~=~
\L_{\chi'} ~=~ k\L - \mu$. The operator is given explicitly by
the same polynomial as in \sinf, i.e.,
\eqn\laca{ _k\ci^\L_\mu ~( 
\underbrace{ \vf \otimes \dots \otimes \vf }_k ) ~~=~~
\widehat{_k\cp^\mu} ~ ( \underbrace{
\vf \otimes \dots \otimes \vf }_k ) }
where the hat on ~$_k\cp^\mu$~ symbolizes the right action
\raca{b}, the explicit action of a typical term of
~$ _k\cp^\mu$~ being (cf. \deco{c})~:}
\eqn\lacat{ \eqalign{ 
&\widehat{
\left\{ X^-_{\beta^1_1} \dots X^-_{\beta^1_{n_1}} ~\otimes ~
\dots ~\otimes ~ 
X^-_{\beta^k_1} \dots X^-_{\beta^k_{n_k}} \right\} } ~
\underbrace{ \vf \otimes \dots \otimes \vf }_k ~~=\cr &\quad =~~
\left( \hat X^-_{\beta^1_1} \dots \hat X^-_{\beta^1_{n_1}} \vf \right) 
~\dots ~ \left( \hat X^-_{\beta^k_1} \dots \hat
X^-_{\beta^k_{n_k}} \vf \right) \cr}} {\it Proof:}~~ Completely
analogous to the case $k=1$ (cf. \Dob). ~\bu

\st
{\it Remark 2:} ~~The analog of the intertwining property \intc\
on the group level, i.e.,
\eqn\intd{ _k\ci^\L_\mu ~\circ~
\underbrace{ \tih (g) \otimes \dots \otimes \tih (g) }_k 
~~=~~ \tihp (g) ~\circ ~ {_k\ci}^\L_\mu ~, \quad \forall g\in G }
will hold, in general, for less values of ~$\L$~ than \intc.
This in sharp contrast with the ~$k=1$~ case, where there is no
difference in this respect. An additional feature on the group
level common for all ~$k\geq 1$~ is that some discrete
representation parameters of ~$\chi$, not represented in ~$\L$,
get fixed.~~\dia

\st
{\it Remark 3:} ~~ Let us stress that since we have realized
arbitrary representations in the spaces of scalar-valued
functions ~$\vf$~ then also the \idos\ are ~{\it scalar}~
operators in all cases $-$ geometrically speaking, these
operators intertwine (tensor products of) line bundles. This
simplicity may be contrasted with the proliferation of tensor
indices in the approaches relying on Weyl's $SO(n)$ polynomial
invariant theory \Wey, cf., e.g., \EG, \BEG, where ~$G ~=~
SO(n+1,1)$, ~$M ~=~ M_0 ~=~ SO(n)$, ~$\dim A ~=~ 1$.~~\dia 

Finally we should mention that the simplest formulae are obtained
of one restricts the functions to the conjugate to $N$ subgroup
~$\tn$~ \Dob~:
\eqn\nonc{ C_\chi ~~\doteq ~~ \{ \phi ~=~ R~ \vf ~\vert ~
\vf\in\tcih \,, ~~ (R \vf) (\nn) ~\doteq ~\vf (\nn)
\,, ~~ \nn\in\tn \} } 
Clearly, the elements of ~$\cih$~ and consequently ~$\tcih$~ are
almost determined by their values on $\tn$ because of right
covariance \fun\ and because, up to a finite number of
submanifolds of strictly lower dimension, every element of $G$
belongs to ~$\tn MAN$. The latter are open dense submanifolds of
~$G$~ of the same dimension forming non-global Bruhat
decompositions of $G$. Connectedly, ~$\tn$~ is an open dense
submanifold of ~$G/P$.

The ER ~$T^\chi$~ acts in this space by:
\eqn\acnc{ \eqalign{ &\left( T^\chi(g)~\phi \right) (\nn)
~~=~~ e^{\nu(H)} ~ D^\mu(m^{-1}) ~\phi (\nn') \cr
& D^\mu(m) ~\phi (\nn) ~~=~~ D^\mu(m) ~\vf (\nn) ~~=~~ 
\lg ~v_0 ~, ~ D^\mu(m) ~\cf(\nn) ~\rg \cr } }
where ~$g\in G$, ~$\nn,\nn'\in\tn$, ~ $m\in M$,
~$a= \exp(H), H\in\ca$, and we have used the Bruhat
decomposition ~$g^{-1}\nn ~=~ \nn'man$, ($n\in N$). [The
transformation can also be defined separately for ~$g^{-1}\nn
\notin \tn MAN$~ and there exists smooth passage from \acnc\ to
these expressions. This is related to the passage between
different coordinate patches of ~$G/P$.] One may easily check
that the restriction operator ~$R$~ intertwines the two
representations, i.e.
\eqn\rint{ T^\chi (g) ~ R ~~=~~ R ~ \tih (g) \,, \quad 
\forall g\in G}

\vskip 5mm

\newsec{\quad Bilinear operators for SL(n,R) and SL(n,C)} 

\newsubsec{} 
In the present Section we restrict ourselves to the case ~$G ~=~
SL(n,\bbr)$, mentioning also which results are extendable to
$SL(n,\bbc)$. We use the following matrix representations for
$G$, its Lie algebra $\cg$ and some subgroups and subalgebras:
\eqna\subg
$$\eqalignno{ &G ~~=~~ SL(n,\bbr) ~~=~~ \{ g\in gl(n,\bbr)
~\vert~ det~g ~=~ 1 \} &\subg a\cr
&\cg ~~=~~ sl(n,\bbr) ~~=~~ \{ X\in gl(n,\bbr) ~\vert~ tr~X ~=~ 0
\} &\subg b\cr
&K ~~=~~ SO(n) ~~=~~ \{ g\in SL(n,\bbr) ~\vert~ g\ g^t ~=~ g^t\ g
~=~ 1_n \} &\subg c\cr 
&\ca_0 ~~=~~ \{ X\in sl(n,\bbr) ~\vert~ X \ {\rm diagonal} \} 
&\subg d\cr 
&A_0 ~~=~~ \exp (\ca_0) ~~=~~ \{ g\in SL(n,\bbr) ~\vert~ g\ {\rm
diagonal} \} &\subg e\cr &\eqalign{ 
M_0 ~~=&~~ \left\{ m\in K ~ \vert ~ ma=am , ~\forall a \in A_0
\right\} ~=\cr 
=&~ \{ m ~=~ {\rm diag}(\d_1,\d_2,...,\d_n) ~\vert~ \d_k
= \pm, \quad \d_1\d_2...\d_n ~=~ 1 \} 
~~=~~ M^d_0 \cr} &\subg f\cr
&\cn_0 ~=~ \{ X = (a_{ij}) \in gl(n,\bbr) ~\vert~ a_{ij} ~=~ 0
~, \quad i\geq j \} 
&\subg g\cr &\eqalign{ N_0 ~=&~ \exp (\cn_0) ~=~ 
\{ X = (a_{ij}) \in gl(n,\bbr) ~\vert~ a_{ii} ~=~ 1, \quad a_{ij}
~=~ 0 ~, \quad i> j \} 
\cr } \qquad &\subg h\cr
&\tcn_0 ~=~ \{ X = (a_{ij}) \in gl(n,\bbr) ~\vert~ a_{ij} ~=~ 0
~, \quad i\leq j \} 
&\subg i\cr &\eqalign{ 
\tn_0 ~=&~ \exp (\tcn_0) ~=~ 
\{ X = (a_{ij}) \in gl(n,\bbr) ~\vert~ a_{ii} ~=~ 1, \quad a_{ij}
~=~ 0 ~, \quad i< j \} 
\cr } \qquad 
&\subg j\cr
}$$

Since the algebra $sl(n,\bbr)$ is maximally split then the Bruhat
decomposition with the minimal parabolic:
\eqn\bru{ \cg ~~=~~ sl(n,\bbr) ~~=~~ \tn_0 ~\oplus ~\ca_0 ~\oplus
~ \tcn_0 } 
may be viewed as a restriction from $\bbc$ to $\bbr$ of the
triangular decomposition of its complexification:
\eqn\gau{ \cg^\bac ~~=~~ sl(n,\bbc) ~~=~~ \cg^\bac_+ ~\oplus ~\ch
~\oplus ~ \cg^\bac_- } 
Accordingly, we may use for both cases the same Chevalley basis
consisting of the $3(n-1)$ generators $X^+_i$, $X^-_i$, $H_i$
given explicitly by:
\eqn\che{ X^+_i ~=~ E_{i,i+1} \ , \qquad X^-_i ~=~ E_{i+1,i}\ ,
\qquad H_i ~=~ E_{ii} - E_{i+1,i+1} \ , \qquad i = 1,...,n-1 }
where $E_{ij}$ are the standard matrices with 1 on the
intersection of the $i$-th row and $j$-th column and zeroes
everywhere else. Note that $X^+_i$, $X^-_i$, $H_i$, resp.,
generate $\tn_0$, $\tcn_0$, $\ca_0$, resp., over $\bbr$ and
$\cg^\bac_+$, $\cg^\bac_-$, $\ch$, resp., over $\bbc$.

\newsubsec{} 
We consider induction from the minimal parabolic case, i.e., $P=
M_0A_0N_0$. The characters of the discrete group ~$M_0=M^d_0$~
are labelled by the signature: ~$\eps ~=~ (
\eps_1,\eps_2,...,\eps_{n-1} )$, ~$\eps_k ~=~ 0,1$~: 
\eqn\cha{ {\rm ch}_\eps (m) ~~=~~ 
{\rm ch}_\eps \left( \d_1,\d_2,...,\d_n \right)
~~ \doteq ~~ \prod_{k=1}^{n-1} ~(\d_k)^{\eps_k} ~, 
\qquad m\in M_0 }
The (nonunitary) characters ~$\nu \in\ca^*_0$~ of ~$A_0$~ are
labelled by ~$c_k\in\bbc$, $k=1,...,n-1$, which is the value of
~$\nu$~ on ~$H_k ~=~$diag$(0,...,0,1,-1,0,...,0) ~\in\ca_0$,
(with the unity on $k$-th place), $k=1,...,n-1$, i.e.,
~$c_k=\nu(H_k)$~:
\eqn\chb{ \eqalign{ {\rm ch}_c(a) ~~=&~~ {\rm ch}_c \left(\exp
\sum_k t_kH_k \right) ~~\doteq~~ \exp \sum_k t_k \nu(H_k) ~~=~~
\exp \sum_k t_kc_k ~~=~~ \prod_k \ha_k^{c_k} \cr a ~~=&~~ \prod_k
a_k ~\in A-0 ~, \quad a_k ~=~ \exp t_k H_k ~\in A_0 ~, \qquad
t_k, \ha_k = \exp t_k \in \bbr } }

Thus, the right covariance property \fun\ is:
\eqn\fub{ \cf(gman) ~~=~~ \prod_{k=1}^{n-1} ~(\d_k)^{\eps_k} ~
\ha_k^{c_k} ~\cf(g) } and in this case we have scalar functions,
i.e., ~$\vf ~=~ \cf$~ because ~$M_0$~ is discrete. Thus, the ER
acts on the restricted functions as (cf. \acnc):
\eqn\trf{\left( T^{c,\eps}(g)~\phi \right) (\nn)
~~=~~ \prod_{k=1}^{n-1} ~(\d_k)^{\eps_k} ~\ha_k^{c_k}
~\phi (\nn') } (note that $\d_k = (\d_k)^{-1}$). The functions
$\phi$ depend on the $n(n-1)/2$ nontrivial elements of the
matrices of $\tcn_0$. For further use those will be denoted by
$z^i_j$, i.e., for $\nn \in\tcn_0$ we have $\nn = (a_{ij})$ with
$a_{ij} = z^i_j$ for $i> j$, cf. \subg{j}.

The correspondence between the ER with signature ~$\chi ~=~
[c,\eps]$~ and the highest weight ~$\L$~, ~ used in the general
construction of the previous Section, here is very simple \Dob:
~$\L ~=~ -\nu$, so that ~$\L(H) ~=~ -\nu(H)$. Further, we recall
that the root system of $sl(n,\bbc)$ is given by roots:
~$\pm\a_{ij}$, $i<j$, so that $\a_{ij} = \a_i +
\a_{i+1} + \dots + \a_{j-1}$ for $i+1 <j$ and $\a_{i,i+1} =
\a_i$, where $\a_i$, $i=1,...,n-1$, are the simple roots with
non-zero scalar products: ~$(\a_i,\a_i) ~=~ \a_i(H_i) ~=~ 2$,
~$(\a_i,\a_{i+1}) ~=~ \a_i(\a_{i+1}) ~=~ -1$, then ~$\a_i^\vee
~=~ \a_i$. Then we use $(\nu,\a_i) ~=~ \nu(H_i) ~=~ c_i$.

We need also the infinitesimal version of \trf:
\eqn\infs{ \tcl^c(Y) ~\phi (\nn) ~~\doteq ~~
\left( {d\over dt} \left( T^{c,\eps}(\exp tY) ~\phi \right) (\nn)
\right)_{\vert_{t=0}} ~, \quad Y \in\cg }
which we give for the Chevalley generators \che\ explicitly:
\eqna\bl $$\eqalignno{ 
& \tcl^c (X^+_i) ~=~ z^{i+1}_{{i}} \left( c_{{i}} +
\sum_{k=i+1}^{{n}} N^{k}_i ~-~ \sum_{k=i+2}^{{n}} N^{k}_{i+1}
\right) ~-~ \sum_{s=1}^{{i-1}} z^{i+1}_s
~D^{{i}}_s ~+~ \sum_{k=i+2}^n z^k_i D^k_{i+1} \qquad \qquad &\bl
a\cr & \tcl^c (X^-_i) ~=~ -~ D^{i+1}_i ~-~ \sum_{s=1}^{i-1}
z^{i}_s D^{i+1}_s &\bl b\cr &\tcl^c (H_{i}) ~=~ c_i ~+~
\sum_{k=i+1}^n N^k_i ~-~ \sum_{k={i+2}}^n N^k_{i+1} ~-~
\sum_{s=1}^{i-1} N^i_s ~+~ \sum_{s=1}^{{i}} N^{i+1}_s &\bl c\cr
}$$ where ~$D^i_j ~\equiv ~{\pd\over \pd z^i_j}$~, ~$N^i_j
~\equiv ~z^i_j\ {\pd\over \pd z^i_j}$~, and we are using the
convention that when the lower summation limit is bigger than the
higher summation limit than the sum is zero.

We also need the right action \rac\ for the lowering generators
which on the restricted functions is given explicitly by:
\eqn\racn{ {\hat X}^-_i\ \phi (\nn) ~~=~~ \left( D^{i+1}_i ~+~
\sum_{k=i+2}^{n} z^{k}_{i+1} D^{k}_i \right) \ \phi (\nn) }

Naturally the signature ~$\eps$~ representing the discrete
subgroup ~$M=M^0_d$~ is not present in \bl{}, \racn. Thus,
formulae \bl{}, \racn are valid also for the holomorphic ERs of
$SL(n,\bbc)$.

Now to obtain explicit examples of multilinear \idos\ it remains
to substitute formula \racn\ in the corresponding formulae for
the singular vectors of the k-Verma modules. We note that often a
singular vector will produce many \idos. For example each formula
valid for any simple root will produce $n-1$ formula, each
formula valid for roots as $\a_1 +\a_2$ will produce $n-2$
formulae for each $\a_i + \a_{i+1}$. To save space we shall not
write these formulae except in a few examples in the cases $n=2$
and $n=3$.

\newsubsec{}
Now we restrict ourselves to the case ~$G ~=~ SL(2,\bbr)$. We
denote: ~$x=z^2_1$, ~$c=c_1$, ~$\eps = \eps_1$, ~$X^\pm =
X^\pm_1$, ~$H = H_1$. We start with ~{\rm bilinear}~ \idos,
$k=2$. We combine Propositions 2 \& 3. For ~$\cg ~=~ sl(2,\bbr)$~
Proposition 2 gives ~{\it all singular vectors}~ of bi-Verma
modules since all weights in ~$\G^+$~ are of the form ~$\mu ~=~
n\a\,$, $n\in\bbn$. Thus we have:

\st
{\bf Theorem 1:} ~~{\it All bilinear \idos\ for the case of ~$\cg
~=~ sl(2,\bbr)$~ are given by the formula:
\eqn\bili{ _2\ci^\L_{n\a} \left( \phi \right) ~~=~~
\sum_{j=0}^{[n/2]} ~\g^\L_{nj}~ \phi^{(n-j)} ~\phi^{(j)} } where
~$\phi^{(p)} ~\doteq ~ (\pd_x)^p \phi (x)$, ~$\pd_x ~\doteq ~
\pd/\pd x\,$, and the coefficients ~$\g^\L_{nj}$~ are given in
Proposition 2. The intertwining property is:}
\eqn\intx{ \eqalign{ &_2\ci^\L_{n\a} ~\circ~
\left( \tcl^c (X) \otimes 1_u ~+~ 1_u \otimes \tcl^c (X) \right)
~~=~~ \tcl^{c'} (X) ~\circ ~ {_2\ci}^\L_{n\a} ~, \quad
\forall X\in \cg \cr &c ~=~ -\L(H), \quad c' ~=~ 2(c+n) \cr} }
{\it Proof:}~~ Elementary combination of Propositions 2 \& 3 in
the present setting. In particular, ~$\L' ~=~ 2\L - n\a$, ~$c'
~=~ -\L'(H) ~=~ -2\L(H) + n\a(H) ~=~ 2(c+n)$.~\bu

As we mentioned, the corresponding intertwining property on the
group level is restricting the values of ~$\L$~ and, as for
$k=1$, of some discrete parameters not represented in ~$\L$. In
the present case we have:

\st
{\bf Theorem 2:} ~~{\it All bilinear \idos\ for the case of ~$G
~=~ SL(2,\bbr)$ are given by formulae \bili\ and \verc\ with
`integer' highest weight: ~$\L(H) ~=~ p \in\bbz$. The
intertwining property is:}
\eqn\intw{ \eqalign{ &_2\ci^\L_{n\a} ~\circ~ \left( T^{c,\eps}
(g) \otimes T^{c,\eps} (g) \right) ~~=~~ T^{c',\eps'} (g) ~\circ
~ {_2\ci}^\L_{n\a} \,, \quad \forall g\in G \cr &c ~=~ -\L(H) ~=~
-p, \quad c' ~=~ 2(c+n) ~=~ 2(n-p) ,
\quad \eps ~=~ \eps' ~=~ p\,({\rm mod}~2) 
\cr} }
{\it Proof:}~~ Follows by using Theorem 1 and checking \intw\ for
~$g ~=~ \pmatrix{ 0 &1 \cr -1 &0}$
\hfil\break which sends ~$x\neq 0$~ into ~$-1/x\,$~\bu \nl 
Thus, the signatures of the two intertwined spaces coincide and
are determined by the parameter ~$p$.

Next we consider the example of ~{\it invariant}~ functions
$\phi$, i.e., functions for which the transformation law \trf\
has no multipliers - this happens iff ~$c ~=~ \eps ~=~ 0$. For
these functions the bilinear \idos\ are given by the special case
of Theorem 2 when ~$\L(H) ~=~ p ~=~ 0$, which by \verc\ further
restricts ~$n$~ to be even or ~$n=1$. Formula \bili\ with \verc\
substituted simplifies to:
\eqn\bilp{ _2\ci^0_{n\a} \left( \phi \right) ~~=~~
\sum_{j=1}^{n/2} ~ (-1)^{j-1} ~(1-\half \d_{j,{n\over 2}}) ~ {n-j
\over n(n-1)} ~ {n \choose j} ~ {n-1 \choose j-1} ~
\phi^{(n-j)} ~\phi^{(j)} \,, \quad n\in 2\bbn } 
\eqn\bilo{ _2\ci^0_{\a} \left( \phi \right) ~~=~~ \phi ~ \pd_x
\phi ~~=~~ \phi ~\phi' } and in addition we have fixed the
constant ~$\g_0$~ for later convenience. Let us write out the
first several cases of \bilp~:
\eqna\bilr
$$\eqalignno{ &_2\ci^0_{2\a} \left( \phi \right) ~~=~~ \half ~
\left( \phi' \right)^2 &\bilr a\cr &_2\ci^0_{4\a} \left( \phi
\right) ~~=~~ \phi'''~\phi' ~-~ \thf ~(\phi'')^2 &\bilr b\cr
&_2\ci^0_{6\a} \left( \phi \right) ~~=~~ \phi^{(5)}~\phi' ~-~ 10
~ \phi^{(4)}~ \phi'' ~+~ 10 ~(\phi''')^2 &\bilr c\cr }$$ where
(standardly) ~$\phi' \equiv \pd_x\phi ~=~ \phi^{(1)}$, ~$\phi''
\equiv \pd^2_x\phi ~=~ \phi^{(2)}$, ~$\phi''' \equiv \pd^3_x\phi
~=~ \phi^{(3)}$. Note that \bilr{b} (i.e., \bilp\ for ~$n=4$)
was already given in \schw. We give now two important technical
statements.

\st
{\bf Lemma 1:} ~~ {\it For ~$n>2$~ the (formal) substitution
~$\phi(x) ~\mt ~ { \a x -\g \over \d - \b x}$~ in the \idos\
\bilp\ gives zero:}
\eqn\zer{ _2\ci^0_{n\a} \left( \phi_0 \right) ~~=~~ 0 \,, \quad
\phi_0(x) ~\equiv ~ { \a x -\g \over \d - \b x}
\,, \quad n \in 2 + 2\bbn } 
{\it Proof:}~~ By direct substitution. In the calculations one
uses the fact:
\eqn\zerd{ \pd_x^m ~ { \a x -\g \over \d - \b x}
~~=~~ { (-1)^m ~ m! ~ \b^{m-1} \over (\d - \b x)^{m+1} }
\,, \quad m\in\bbn }
After the substitution of \zerd\ in \bilp\ the resulting
expression is proportional to ~$(1 ~-~ 1)^{n-2}$~ which is zero
for ~$n>2$; (the latter making clear why the Lemma is not valid
for $n=2$).~\bu

\st
{\bf Lemma 2:} ~~ {\it Let ~$\phi,\psi\in$~Diff$_0S^1$, the group
of orientation preserving diffeomorphisms of the circle ~$S^1 ~=~
\bbr/2\pi\bbz$. Then we have:}
\eqna\difp \eqna\difo
$$\eqalignno{ &_2\ci^0_{n\a} \left( \phi\circ\psi \right) ~~=~~
\left( \psi'\right)^n ~ {_2\ci}^0_{n\a} \left( \phi \right)
\,, \quad n=1,2 &\difp {}\cr 
&\eqalign{ _2\ci^0_{n\a} \left( \phi\circ\psi \right) ~~=&~~
\left( \psi'\right)^n ~ _2\ci^0_{n\a} \left( \phi \right) ~+~
\left(\phi'\right)^2 ~ {_2\ci}^0_{n\a} \left( \psi \right) 
~+\cr &+~ P_n(\phi,\psi) \,, \quad n\in 2 +2\bbn \cr} &\difo a\cr
&P_4(\phi,\psi) ~~=~~ 0 &\difo b\cr &P_n(\phi,\phi_0) ~~=~~ 0
\,, \quad n\in 4 +2\bbn &\difo c\cr 
}$$ {\it Proof:}~~ For \difp{} and \difo{a,b} this is just
substitution. Further we note that:
\eqn\ner{_2\ci^0_{n\a} \left( \phi\circ\phi_0 \right)
~~=~~ \left( \phi'_0\right)^n ~ {_2\ci}^0_{n\a} \left( \phi
\right)\,, \quad \phi_0(x) ~=~ { \a x -\g \over \d - \b x} } is
just the intertwining property of ~$_2\ci^0_{n\a}$~ and then
\difo{c} follows because of Lemma 1.~\bu

\st
{\it Remark 4:}~~ We give an example from the last Lemma:
\eqn\pes{ \eqalign{ P_6(\phi,\psi) ~~=&~~
10 ~ \left( \psi' \right)^2 ~\left( 3 ~ \phi'''~ \phi' ~-~ 4
~\left( \phi'' \right)^2 \right) ~{_2}\ci^0_{4\a} \left( \psi
\right) ~-\cr &-~ 5~ \phi'' ~\phi' ~\left( \psi^{(4)} ~ \left(
\psi'\right)^2 ~-~ 6 ~\psi''' ~\psi'' ~\psi' ~+~ 6 ~\left(
\psi''\right)^3 \right) \cr}} Using \zerd\ it is straightforward
to show ~$P_6(\phi,\phi_0) ~=~ 0\,$. In fact, the first term in
\pes\ vanishes for ~$\psi ~=~ \phi_0$~ because of \zer. The
vanishing of the second term in \pes\ prompts us that the
trilinear expression in ~$\psi$~ is also a \ido. This is indeed
so, cf. the last Section for some more examples for trilinear
operators.~~\dia

We can introduce now a hierarchy of ~$GL(2,\bbr)$~ invariant
~$\han$-differentials for every ~$n\in 2\bbn$~:
\eqna\schwc
$$\eqalignno{ &{\rm Sch}_n (\phi) ~~\doteq~~ {_2}\ci^0_{n\a}
\left( \phi \right) ~ \left( { dx \over \phi'} \right)^{n/2}
\,, \quad n\in 2\bbn &\schwc a\cr &{\rm Sch}_n \left( \phi \circ
\phi_0 \right) ~~=~~ {\rm Sch}_n \left(\phi\right) ~\circ ~
\phi_0 \,, \quad \phi_0(x) ~=~ { \a x -\g \over \d - \b x}
&\schwc b\cr }$$ where the property \schwc{b} is just a
restatement of \ner\ for $n>2$ and \difp{} for $n=2$. The usual
Schwarzian ~Sch$_4$~ is one of these objects (cf. \schwa). It has
an additional property~:
\eqn\prop{ {\rm Sch}_4 \left( \phi \circ \psi \right) ~~=~~ {\rm
Sch}_4 \left(\phi\right) ~\circ ~ \psi ~+~ {\rm Sch}_4
\left(\psi\right) \,, \quad \phi, \psi \in {\rm Diff}_0S^1 }
showing that it is a 1-cocycle on Diff$_0S^1$ \Kib, \Kia.

\st
{\it Remark 5:} ~~One may consider also ~{\it
half-differentials}~ and using \schwc{a} and \bilo\ write:
\eqn\had{ {\rm Sch}_1 (\phi) ~~\doteq~~ {_2}\ci^0_{\a} \left(
\phi \right) ~
\left( { dx \over \phi'} \right)^{1/2} ~~=~~
\phi ~ \left( \phi' dx \right)^{1/2} ~~=~~
\phi ~ \left( d \phi \right)^{1/2} } Property \schwc{b} then
follows from \difp{}.~~\dia

Finally, we just mention the case when the resulting functions
are invariant: ~$c'=0$ ~$\Lra$~ $c = -n$. This is only possible
when ~$n$~ is even, cf. \verc. Formula \bili\ with \verc\
substituted simplifies considerably:
\eqn\bils{ \eqalign{ _2\ci^{-n}_{n\a} \left( \phi \right) 
~~=&~~ \sum_{j=0}^{n/2} ~ (-1)^j ~(1-\half\d_{j,n/2}) ~ 
\phi^{(n-j)} ~\phi^{(j)} ~=\cr =&~~ \phi^{(n)} ~\phi ~-~
\phi^{(n-1)} ~\phi' ~+~ \phi^{(n-2)} ~\phi'' ~-\cr &-~
\phi^{(n-3)} ~\phi''' ~+ \ldots +~ \half ~(-1)^{n/2} ~\left(
\phi^{(n/2)} \right)^2 \,, \quad n\in 2\bbn }}
and we have fixed the constant ~$\g_0$~ appropriately. 

\newsubsec{}
Now we consider to the case ~$G ~=~ SL(3,\bbr)$. We denote:
~$x=z^2_1$, ~$y=z^3_2$, ~$z=z^3_1$. The right action is ($\phi
~=~ \phi (x,y,z)$)~:
\eqn\racd{ {\hat X}^-_1\ \phi ~~=~~ \left( \pd_x ~+~ y\pd_z
\right) \ \phi ~, \qquad {\hat X}^-_2\ \phi 
 ~~=~~ \pd_y\ \phi ~, \qquad {\hat X}^-_3\ \phi ~~=~~ \pd_z\ \phi
} and we have given it also for the nonsimple root vector $X^-_3
= [X^-_2,X^-_1]$.

The bilinear operator correponding to \sngss\ is:
\eqn\sngsd{ \eqalign{ _2^3\ci^\L_{\a} (\phi) ~~=&~~ \phi\ \left(
(\L_1 -\L_2) \ \left( \phi_{xy} ~+~ y \phi_{yz} \right) 
~-~ \L_2~ \phi_z \right) \ ~-\cr &-~ ( \L_1 + \L_2 +1 )\ \phi_y\
(\phi_x ~+~ y\phi_z) }} where $\a = \a_3 = \a_1 +\a_2$, $\phi_x
\equiv {\pd\phi\over \pd_x}$, etc. The two bilinear operators
correponding to \sbi{} are given by:
\eqn\sbd{ \eqalign{ _2^3\ci^\L_{2\a} (\phi) ~=&~ (a_1 \ +\ a_2 \
+\ 2a_3) \ \phi \ \phi_{zz} ~+~ (a_2 \ +\ 2a_3) \ \phi \
\phi_{xyz} ~+~ (a_2 \ +\ 4a_3)\ y \ \phi\ \phi_{yzz} ~+\cr &+~
a_3 \ \phi \ \left( \phi_{xxyy} \ +\ y\
\ \phi_{xyyz} \ +\ y^2\ \ \phi_{yyzz} \right) ~+ 
\cr &+~ (b_1 \ + \ 2c_1)\ \phi_{y} \ \left( \phi_{xz} \ +\ y\
\phi_{zz} \right) ~+~ ( b_2\ + \ 2c_2)
\ \left( \phi_{x} \ +\ y\ \phi_{z} \right) \ \phi_{yz} 
\ + \cr &+~ c_1 \ \phi_{y} \ \left( \phi_{xxy} \ +\ 2y\
\phi_{xyz} \ +\ y^2\ \phi_{yzz} \right) ~+\cr &+~ c_2 \ \left(
\phi_{x} \ +\ y\ \phi_{z} \right) \
\left( \phi_{xyy} \ +\ y\ \phi_{yyz} \right) 
~+ \cr &+~ (d_1\ + \ d_2 \ + \ d_3) \ \phi^2_{z} 
~+~ ( d_2\ + \ 2d_3) \ \phi_{z} 
\ \left( \phi_{xy} \ +\ y\ \phi_{yz} \right) 
~+ \cr &+~ d_3 \ \left( \phi_{xy} \ +\ y\ \phi_{yz} \right)^2 ~+~
d_4 \ \phi^2_{y}
\ \left( \phi_{x} \ +\ y\ \phi_{z} \right)^2 
\cr } } 
with constants as given in \sbi{b,c}.

The intertwining property is:
\eqn\intx{ _2^3\ci^\L_{n\a} ~\circ~
\left( \tcl^c (X) \otimes 1_u ~+~ 1_u \otimes \tcl^c (X) \right)
~~=~~ \tcl^{c'} (X) ~\circ ~ {_2^3\ci}^\L_{n\a} ~, \quad
\forall X\in \cg }
where ~$c_i ~=~ -\L_i ~=~ -\L(H_i)$, ~$\L' ~=~ 2\L - n\a$, ~$c'_i
~=~ -\L'(H_i) ~=~ -2\L(H_i) + n\a(H_i) ~=~ -2\L_i + n ~=~
2c_i+n$, ~$i=1,2$, since $\a(H_i) = (\a_1+\a_2)(H_i) = 1$.

The case of invariant functions, i.e., $c_k=\eps_k=0$ gives a
trivial (zero) operator $n=2$, while for $n=1$ we have (up to
scalar multiple):
\eqn\sngsz{ _2^3\ci^0_{\a} (\phi) ~~=~~ \phi_y\ (\phi_x ~+~
y\phi_z) } 

In the case of invariant resulting functions, i.e.,
$c'_k=\eps'_k=0$, ~$\L ~=~ n\a/2$, ~$\L_i ~=~ n/2$, we have from
\sngsd:
\eqn\sngsdd{ _2^3\ci^{\a/2}_{\a} (\phi) ~~=~~ 
 \half\ \phi_z ~+~ 2\ \phi_y\ (\phi_x ~+~ y\phi_z) }
while from \sbd\ we have two operators corresponding to the two
solutions given by \sbi{b,c}, resp.:
\eqna\sbdd 
$$\eqalignno{ _2^3\ci^\a_{2\a} (\phi) ~~=&~~ 2\ \phi \ \phi_{zz}
~-~ 2\ y \ \phi\ \phi_{yzz} ~-&\cr &-~ \phi \ \left( \phi_{xxyy}
\ +\ y\ \ \phi_{xyyz} \ +\ y^2\ \ \phi_{yyzz} \right) ~- &\cr &-~
2\ \phi_{y} \ \left( \phi_{xz} \ +\ y\ \phi_{zz} \right) ~+~ 6 \
\left( \phi_{x} \ +\ y\ \phi_{z} \right) \ \phi_{yz}
\ + &\cr &+~ 2 \ \phi_{y} \ \left( \phi_{xxy} \ +\ 2y\ \phi_{xyz}
\ +\ y^2\ \phi_{yzz} \right) ~+&\cr &+~ 2 \ \left( \phi_{x} \ +\
y\ \phi_{z} \right) \ \left( \phi_{xyy} \ +\ y\ \phi_{yyz} \right) 
~- &\cr &-~ 2 \ \phi^2_{z} ~-~ 2 \ \phi_{z} \ \left( \phi_{xy} \
+\ y\ \phi_{yz} \right) ~-~ 2 \ \left( \phi_{xy} \ +\ y\
\phi_{yz} \right)^2 &\sbdd b\cr &&\cr _2^3\ci'^\a_{2\a} (\phi)
~~=&~~ \phi^2_{y}\ \left( \phi_{x} \ +\ y\ \phi_{z} \right)^2 
&\sbdd a\cr }$$

\vskip 5mm

\newsec{\quad Examples with ~k ~$\geq$ ~3}

We return now to the ~$GL(2,\bbr)$~ setting to give examples of
trilinear \idos\ using the singular vectors of ~tri-Verma modules
above. The trilinear \idos\ for the case of ~$\cg ~=~
sl(2,\bbr)$~ are given by the formula:
\eqn\bili{ _3\ci^\L_{n\a} \left( \phi \right) ~~=~~ \sum_{{ j,k
\in\bbz_+ \atop n-j-k \geq j \geq k} } ~
\g^\L_{njk}~ \phi^{(n-j-k)} ~\phi^{(j)} ~\phi^{(k)} } 
where the coefficients ~$\g^\L_{nj}$~ are given from the
expressions for the corresponding singular vectors of ~tri-Verma
modules, e.g., those given in the previous subsection.

If we pass to the group level then the possible weights are
restricted to be 'integer' (as in Theorem 2)~: ~$\L(H) ~=~ p~
\in ~\bbz$~ and the corresponding intertwining property
is~:
\eqn\intwr{ \eqalign{ &_3\ci^\L_{n\a} ~\circ~ \left( T^{c,\eps}
(g) \otimes T^{c,\eps} (g) \otimes T^{c,\eps} (g) \right) ~~=~~
T^{c',\eps'} (g) ~\circ ~ {_3\ci}^\L_{n\a}
\,, \quad \forall g\in G \cr
&c ~=~ -\L(H) ~=~ -p, \quad c' ~=~ 3(c+n) ~=~ 3(n-p) ,
\quad \eps ~=~ \eps' ~=~ p\,({\rm mod}~2) 
\cr} } 

Next we restrict to the example of ~{\it invariant}~ functions
$\phi$, i.e., ~$c ~=~ \eps ~=~ 0$. The trilinear \idos\ obtained
from the singular vectors in \sngr{} are:
\eqna\intr
$$\eqalignno{ &_3\ci^0_{\a} \left( \phi \right) ~~=~~ \left( \phi
\right)^2 ~\phi' &\intr a\cr &_3\ci^0_{2\a} \left( \phi \right)
~~=~~ \phi ~ \left( \phi' \right)^2 &\intr b\cr &_3\ci^0_{3\a}
\left( \phi \right) ~~=~~ \left( \phi' \right)^3 &\intr c\cr
&_3\ci^0_{4\a} \left( \phi \right) ~~=~~ \phi ~ \left(
\phi'''~\phi' ~-~ \thf ~(\phi'')^2 \right) &\intr d\cr
&_3\ci^0_{5\a} \left( \phi \right) ~~=~~ \phi' ~ \left(
\phi'''~\phi' ~-~ \thf ~(\phi'')^2 \right) &\intr e\cr 
&_3\ci^0_{6\a} \left( \phi \right) ~~=~~ \phi^{(4)}~ \left( \phi'
\right)^2 ~-~ 6 ~ \phi''' ~ \phi''~ \phi' ~+~ 6 ~(\phi'')^3
&\intr f\cr &_3\ci'^0_{6\a} \left( \phi \right) ~~=~~ \phi
~\left( \phi^{(5)}~\phi' ~-~ 10 ~ \phi^{(4)}~ \phi'' ~+~ 10
~(\phi''')^2 \right) &\intr {f'}\cr }$$
We recall that the operator in \intr{f} has appeared in \pes. 

Analogously to Lemma 1 we note that for ~$n>3$~ the (formal)
substitution ~$\phi(x) ~\mt ~ { \a x -\g \over \d - \b x}$~ in
the \idos\ \intr{} gives zero:
\eqn\zerr{ _3\ci^0_{n\a} \left( \phi_0 \right) ~~=~~ 0 \,, \quad
\phi_0(x) ~\equiv ~ { \a x -\g \over \d - \b x}
\,, \quad n >3 }
which because of the factorization follows from Lemma 1 except
for \intr{f}.

Analogously to Lemma 2 for ~$\phi,\psi\in$~Diff$_0S^1$~ one can
check for the examples in \intr{}~:
\eqna\difpr \eqna\difor 
$$\eqalignno{ &_3\ci^0_{n\a} \left( \phi\circ\psi \right) ~~=~~
\left( \psi'\right)^n ~ {_3\ci}^0_{n\a} \left( \phi \right)
\,, \quad n=1,2,3 &\difpr {}\cr 
&_3\ci^0_{5\a} \left( \phi\circ\psi \right) ~~=~~ \left(
\psi'\right)^5 ~ _3\ci^0_{5\a} \left( \phi\right) ~+~
\left(\phi'\right)^3 ~ {_3\ci}^0_{5\a} \left( \psi \right) 
&\difor a\cr &_3\ci^0_{6\a} \left( \phi\circ\psi \right) ~~=~~
\left( \psi'\right)^6 ~ _3\ci^0_{6\a} \left( \phi\right) ~+~
\left(\phi'\right)^3 ~ {_3\ci}^0_{6\a} \left( \psi \right) 
~-~ 2 ~\phi'' ~\left(\phi'\psi'\right)^2 ~ _2\ci^0_{4\a}
\left( \psi \right) \qquad &\difor b\cr }$$

Consider now the case of resulting invariant functions, i.e.,
~$c'=\eps'=0$, i.e., ~$p = \L(H) = \hL = n$. There is no operator
for $n=1$, while for $n>1$ we get from \sngr{}:
\eqna\sngd
$$\eqalignno{ _3v_s^{2\a} ~~=&~~ 2 ~\phi'' ~\phi^2 ~-~ \phi'^2
~\phi ~, \qquad \hL = 2 &\sngd b\cr}$$ $$\eqalignno{ _3v_s^{3\a}
~~=&~~ 9 ~ \phi''' ~\phi^2 ~-~ 9 ~\phi'' ~ \phi' ~ \phi ~ ~+~ 4
~\phi'^3 ~ ~, \qquad \hL = 3 &\sngd c\cr}$$ $$\eqalignno{
_3v_s^{4\a} ~~=&~~ 2 ~ \phi^{(4)} ~\phi^2 ~-~ 2 ~\phi''' ~ \phi'
~\phi ~+~ (\phi'')^2 ~ \phi ~ ~, \qquad \hL = 4 &\sngd d\cr}$$
$$\eqalignno{ _3v_s^{5\a} ~~=&~~ 25 ~\phi^{(5)}~ \phi^2 ~-~ 25
~\phi^{(4)} ~ \phi' ~\phi ~+~ 5 ~\phi''' ~ \phi'' ~ \phi ~+&\cr
&~+ ~ 16~ \phi''' ~ \phi'^2 ~-~ 9 ~ (\phi'')^2 ~\phi' ~, \qquad
\hL = 5 &\sngd e\cr}$$ $$\eqalignno{ _3v_s^{6\a} ~~=&~~ 60 ~
\phi^{(4)} ~ \phi'' ~\phi ~-~ 50 ~\phi^{(4)} ~ \phi'^2 ~-~ 45
~\phi'''^2 ~\phi ~+&\cr &~+~ 60 ~\phi''' ~ \phi'' ~ \phi' ~-~ 24
~(\phi'')^3 ~, \qquad \hL = 6 &\sngd f\cr }$$ $$\eqalignno{
_3v'^{6\a}_s ~~=&~~ 2 ~\phi^{(6)} ~\phi^2 ~-~ 2 ~ \phi^{(5)} ~
\phi' ~\phi ~ +~ 2 ~ \phi^{(4)} ~ \phi'' ~\phi ~ -~
\phi'''^2 ~\phi~, \qquad \hL = 6 
&\sngd {f'}\cr }$$

We see that at lower levels there occur many factorizations and
trilinear operators are actually determined by bilinear ones. We
illustrate this by two statements for arbitrary ~k-Verma modules
and the corresponding multilinear \idos.

\st
{\bf Proposition 4:}~~ {\it The singular vectors of the ~k-Verma
modules ~$_kV^\L$ of level ~$n\a$~ with ~$n\in\bbn$, $n\leq k$,
~$\a \in \D_S$~, ~in the case ~$\L(H_\a) ~=~ 0$ are given by:} 
\eqn\sngz{ _kv_s^{n\a} ~~=~~ \g_0 \left\{
\underbrace{ X^-_\a ~\otimes ~ \cdots ~ \otimes ~ X^-_\a}_n ~\otimes 
~\underbrace{ 1_u ~\otimes ~ \cdots ~ \otimes ~ 1_u}_{k-n}
\right\} ~\ho ~v_0 \,, \quad 1 \leq n \leq k\,, \quad
\L(H_\a) = 0}
{\it Proof:}~~ By direct verification.~\bu

\st 
{\bf Proposition 5:}~~ {\it The ~$GL(2,\bbr)$~ multilinear \idos\
with the property:
\eqn\intwrr{ \eqalign{ &_k\ci^0_{n\a} ~\circ~ \underbrace{
T^{c,\eps} (g) \otimes \cdots \otimes T^{c,\eps} (g) }_k ~~=~~
T^{c',\eps'} (g) ~\circ ~ {_k\ci}^0_{n\a}
\,, \quad \forall g\in G \cr
&c ~=~ -\L(H) ~=~ 0, \quad c' ~=~ kn, 
\quad \eps ~=~ \eps' ~=~ 0 \cr} } are given by:}
\eqn\intrr{ _k\ci^0_{n\a} \left( \phi \right) ~~=~~ \phi^{n-k}
~\left( \phi' \right)^n \,,
\quad 1 \leq n \leq k \,, \quad
\L(H) = 0 }
{\it Proof:}~~ Follows from Propositions 3 \& 4.~\bu

We note that the operators in \bilo, \bilr{a}, \intr{a,b,c} are
partial cases of \intrr.

\vskip 5mm

\appendix{A}{Tensor, symmetric and universal enveloping algebras}

Let ~$E$~ be a vector space over ~$F$. The {\it tensor algebra}
~$T(E)$~ over ~$E$~ is defined as the free algebra generated by
the unit element. We have \eqn\tee{ T(E) ~=~
\mathop{\oplus}\limits^\infty_{k=0} T_k(E)~, \quad T_k(E)
~\equiv~ \underbrace{E\otimes \dots \otimes E}_k~, \quad T_0(E)
~=~ F.1 ~. } The elements ~$t\in T_k(E)$~ are called covariant
tensors of rank $k$
\eqn\cov{ t ~=~ \sum t^{i_1\dots i_k} e_{i_1}\otimes \dots
\otimes e_{i_k} }
where ~$e_i\in S$, ~$S$~ is a basis of ~$E$, ~$t^{i_s\dots
i_k}\in F$. (The rank of a covariant tensor does not depend on
the choice of basis of $E$.) The tensor ~$t$~ is called symmetric
tensor if ~$t^{i_n\dots i_k}$~ is symmetric in all indices. Let
us denote
\eqn\sym{ S(E) ~=~ \mathop{\oplus}\limits^\infty_{k=0} S_k(E) }
where ~$S_k(E)$~ is the subspace of all symmetric tensors of rank
$k$. Note that if $\dim E = n <\infty$, then:
\eqn\dimsym{ \dim S_k(E) ~=~ \biggl({n+k-1\atop k}\biggr) }

Let ~$I(E)$, resp., $I_k(E)$, ~ be the two--sided ideal of
~$T(E)$, resp., $T_k(E)$, ~ generated by all elements of the type
~$x\otimes y- y\otimes x,\ x,y\in E$. Then we have
\eqn\teid{ T(E) ~=~ S(E)\oplus I(E) ~, \quad S(E) ~\cong ~T(E)/I(E)
~, \quad S_k(E) ~\cong ~T_k(E)/I_k(E) }

Consider now a Lie algebra ~$\cg$. The universal enveloping
algebra ~$U(\cg)$~ of ~$\cg$~ is defined as the associative
algebra with generators ~$e_i$, where ~$e_i$~ forms a basis of
~$\cg$~ and the relations
\eqn\basl{ e_i\otimes e_j - e_j\otimes e_i ~=~ \sum_k c^k_{ij}e_k
~\equiv~ [e_i, e_j] } hold, where ~$c^k_{ij}$~ are the structure
constants of ~$\cg$. Equivalently ~$U(\cg) \cong T(\cg)/J(\cg)$,
where ~$T(\cg)$~ is the tensor algebra over ~$\cg$, ~$J(\cg)$~ is
the ideal generated by the elements ~$[x,y] - (x\otimes
y-y\otimes x)$. Since ~$T(\cg) = U(\cg) \oplus J(\cg) =
S(\cg)\oplus I(\cg)$~ and ~$J(\cg) \cong I(\cg)$~ are isomorphic,
then also ~$U(\cg) \cong S(\cg)$~ as vector spaces. This is the
content of the Poincar\'e--Birkhoff--Witt (PBW) theorem.

Further in the case of ~$U(\cg)$~ we shall omit the $\otimes$
signs in the expressions for its elements. With this convention
as a consequence of the PBW theorem ~$U(\cg)$~ has the following
basis:
\eqn\basu{e_0 = 1, \ \ e_{i_1\dots i_k} = e_{i_1} \dots 
e_{i_k}, \ \ i_1\leq \dots \leq i_k } 
where we are assuming some ordering of the basis of $\cg$, e.g.,
the lexicographic one.

Finally we recall that ~$\cg$~ and ~$U(\cg)$~ are completely
determined from the following commutation relations:
\eqna\sera
$$\eqalignno{ &[H_i~, ~H_j] ~=~ 0 , ~~~[H_i~, ~X^\pm_j] ~=~ \pm
a_{ij} X^\pm_j &\sera a\cr &[X^+_i~, ~X^-_j] ~=~ \d_{ij} H_i
&\sera b\cr }$$ and Serre relations:
\eqna\sera
$$\eqalignno{ &\sum_{k=0}^n ~(-1)^k ~\left({n \atop k}\right)
~\left(X^\pm_i\right)^k ~X^\pm_j ~\left(X^\pm_i\right)^{n-k} ~=~
0 ~, ~~i \neq j ~, ~~ n = 1 - a_{ij} &\sera c\cr}$$ where
~$X^\pm_i ~, ~H_i ~, ~i=1,\dots ,\ell =$ rank~$\cg$~ are the
Chevalley generators of ~$\cg$, (corresponding to the simple
roots ~$\a_i$), ~$(a_{ij}) = (2(\a_i ~, ~\a_j)/(\a_i ~, \a_i))$~
is the Cartan matrix of ~$\cg$, and ~$(\cdot ~, ~\cdot)$~ is
normalized so that for the short roots ~$\a$~ we have ~$(\a,\a) =
2$.

\vskip 5mm

\newsec{Note Added}

\nt
This note should be read after Remark 4. 
It was published first in our paper:
~~Phys. Atom. Nucl. {\bf 61} (1998) 1735-1742.

\nt
Let us introduce the following notation:
\eqn\nsch{\widetilde{\rm Sch}_n (\phi)\ \doteq\   { 1\over
\phi'^{2} }\ {_2}\ci^0_{n\a}\left(\phi\right) ~~.} 
It is well known (cf., e.g., \KNV) that the famous KdV equation may
be rewritten in the Krichever-Novikov form: 
\eqn\kdvn{\pd_t f\ +\ \widetilde{\rm Sch}_4 (f)\ f'\ \ =\ \  0,\qquad 
f\ =\ f(t,x) ~, }
noting that\ $\widetilde{\rm Sch}_4 (f)$\ is the Schwarz
derivative\ $S[f]$, cf. \schwa. To pass to the standard KdV form:
\eqn\kdvs{\pd_t u\ +\  u'''\ -\  6uu'\ \ =\ \  0 ~,} 
one uses the substitution ~$u\ =\ -\half\widetilde{\rm Sch}_4
(f)$, \KNV. Motivated by the above we make the ~{\bf Conjecture}~ 
that the following equations:
\eqn\conj{\pd_t f\ +\ \widetilde{\rm Sch}_n (f)\ f'\ \ =\ \  0 
\ ,\qquad n\in 2\bbn +2 ~, }
are integrable (true for $n=4$). It may happen (if the
conjecture is true) that this hierarchy of equations coincides
with the KdV one. Then formulae \conj, \nsch\ and \bilp\ would
give an explicit expression for the whole KdV hierarchy in the
Krichever-Novikov form.

\vskip 10mm

\noindent 
{\bf Acknowledgments}
\vskip 0.3cm

\ni 
The author would like to thank H.-D. Doebner for stimulating
discussions and W. Scherer for pointing out ref. \Kia. The
author was supported in part by BNFR under contracts Ph-401 and
Ph-643.

\np 

\parskip=0pt

\listrefs 

\np\end